\documentstyle[epsfig]{mn}

\newif\ifAMStwofonts

   \title{Global velocity field and bubbles in the BCD Mrk86}

   \author[A. Gil de Paz {\rm et al.}]{A. Gil de Paz,
J. Zamorano and J. Gallego
   \\Dept. de Astrof\'{\i}sica, Universidad Complutense de Madrid, 28040 Madrid (Spain)\\
              e-mail: gil@astrax.fis.ucm.es (AGdP)
             }

\date{Accepted 1999 March 1; in original form 1998 May 18}
\pagerange{\pageref{firstpage}--\pageref{lastpage}}
\pubyear{1998}

\begin{document}

\maketitle

\label{firstpage}

\begin{abstract}
   We have studied the velocity field of the Blue Compact Dwarf galaxy
   Mrk86 (NGC~2537) using data provided by 14 long-slit optical
   spectra obtained in 10 different orientations and positions. This
   kinematical information is complemented with narrow-band
   ([OIII]5007\AA\ and H$\alpha$) and broad-band ($B$, $V$, Gunn-$r$
   and $K$) imaging. The analysis of the galaxy global velocity field
   suggests that the ionized gas could be distributed in a rotating
   inclined disk, with projected central angular velocity of
   $\Omega=34$ km\,s$^{-1}$\,kpc$^{-1}$. The comparison between the
   stellar, HI and modeled dark matter density profile, indicates that
   the total mass within its optical radius is dominated by the
   stellar component. Peculiarities observed in its velocity field can
   be explained by irregularities in the ionized gas distribution or
   local motions induced by star formation.

   Kinematical evidences for two expanding bubbles, Mrk86--B and
   Mrk86--C, are given. They show expanding velocities of
   34\,km\,s$^{-1}$ and 17\,km\,s$^{-1}$, H$\alpha$ luminosities of
   3$\times10^{38}$\,erg\,s$^{-1}$ and
   1.7$\times10^{39}$\,erg\,s$^{-1}$, and physical radii of 374 and
   120\,pc, respectively. The change in the [SII]/H$\alpha$,
   [NII]/H$\alpha$, [OII]/[OIII] and [OIII]/H$\beta$ line ratios with
   the distance to the bubble precursor suggests a diminution in the
   ionization parameter and, in the case of Mrk86--B, an enhancement
   of the shock-excited gas emission. The optical-near-infrared
   colours of the bubble precursors are characteristic of low
   metallicity star forming regions ($\sim$0.2\,Z$_{\sun}$) with burst
   strengths of about 1 per cent in mass.

\end{abstract}

\begin{keywords}
galaxies: irregular -- galaxies: compact -- galaxies: individual:
Mrk86 -- galaxies: kinematics and dynamics
\end{keywords}

\section{Introduction}
\label{introduction} 
   The study of the global velocity field of Blue Compact Dwarf
   galaxies (BCD hereafter; Thuan \& Martin 1981) provides important
   clues about their gravitational potential, since these systems are
   rotationally supported (e.g. van Zee et al$.$ 1998).
    
   High spatial resolution HI observations have shown that the
   rotation curves of Blue Compact Dwarf galaxies (Meurer,
   Staveley-Smith \& Killeen 1998; van Zee et al$.$ 1998) and dwarf
   irregulars (dI hereafter; Moore 1994; Flores \& Primack 1994) are
   nearly flat in the galaxy outer regions and have nearly constant
   velocity gradients within their optical radius. Also, optical
   studies of the velocity field of the ionized gas in BCDs obtain
   constant velocity gradients, characteristic of a {\it solid-body}
   rotation law (see Petrosian et al$.$ 1997 for I~Zw~18).

   Although the neutral (see van Zee et al$.$ 1998 and references
   therein) and molecular hydrogen (Young \& Knezek 1989; Israel,
   Tacconi \& Bass 1998) are quite abundant in Blue Compact Dwarfs and
   dwarf irregulars, they are not enough to reproduce the flattening
   of the rotation curve. Like in spiral galaxies, the existence of
   this flattening in the rotation curve of dwarf galaxies has been
   related with the presence of large amounts of dark matter in galaxy
   outer regions (Carignan \& Freeman 1988; Carignan \& Beaulieu 1989;
   Broeils 1992). The dark matter content derived indicates that dark
   matter is even more abundant in dwarfs that in more massive
   galaxies (see Moore 1994 and references therein). In fact, standard
   cold dark matter (CDM hereafter) models predict that low-mass halos
   are denser than more massive systems, because their higher
   formation redshift (Navarro, Frenk \& White 1997, NFW
   hereafter). The density profiles of the simulated CDM halos fall
   with radius as $r^{-2}$. This is density profile expected for a
   flat rotation curve body.
   
   The competition between the dark matter and the stellar mass
   components within the optical radius difficults the analysis of
   {\it solid-body} portion of the rotation curve. Several works have
   argued that dark matter in dwarf galaxies dominates the total mass
   density profile even within their optical radius (Carignan \&
   Beaulieu 1989; Broeils 1992), showing a constant density dark
   matter {\it core} (Moore 1994; Flores \& Primack 1994; Salucci \&
   Persic 1997). On the other hand, Lo, Sargent \& Young
   \shortcite{lo} and Staveley-Smith, Davies \& Kinman \shortcite{ss}
   deduced reasonable virial mass to blue light ratios,
   lower than 7\,M$_{\sun}$/L$_{\sun}$, for a
   large fraction of their samples. Loose \& Thuan \shortcite{loose86}
   found that the virial mass of Haro~2 can be reproduced just adding
   the stellar and HI mass components. Also the study of Swaters
   \shortcite{swaters} of the rotation curves of 44 dwarf galaxies
   indicates that the mass of a large fraction of these galaxies could
   be dominated by the stellar component, even at distances larger
   than three disk scale lengths.

   One of the main sources of uncertainty in all these studies is the
   mass-to-light ratio adopted for the stellar component (Meurer,
   Staveley-Smith \& Killeen 1998; Swarters 1998). Therefore, high
   quality optical and near-infrared imaging and spectroscopy in order
   to obtain the physical parameters of the stellar populations and
   derive reasonable mass-to-light ratios is mandatory to prevent this
   inconvenient which is inherent to this kind of kinematical studies.

\begin{figure}
   \epsfxsize=80mm
   \epsfysize=80mm
   \epsffile{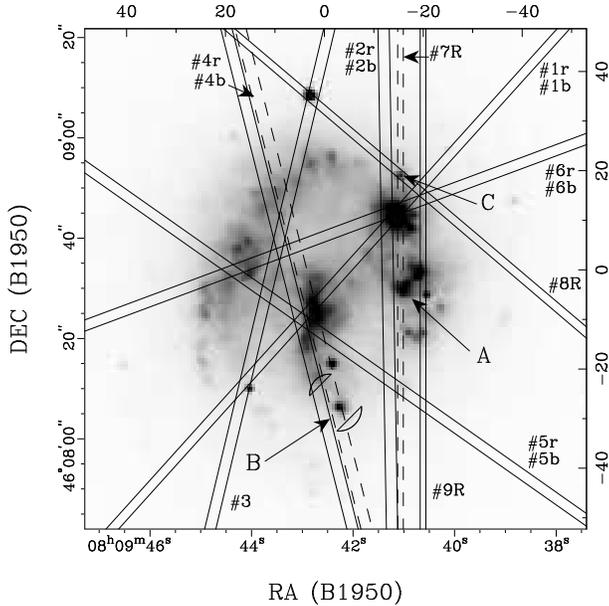}
\caption{Long-slit positions superimposed on the Mrk86 
$B$-band image. Close to the region B, size and position of the Mrk86--B 
bubble H$\alpha$ lobes are shown (see Figures~\ref{bubbleB}a 
\&~\ref{bubbleB}b). Relative coordinates (in arcsec) are referred to the 
$r$-band outer isophotes centre.}

\label{image}
\end{figure}

   Superimposed on the regular {\it solid-body} portion of the
   velocity field, peculiar motions of the ionized gas have been
   observed in many star forming dwarf galaxies (Tomita et al$.$ 1997;
   Petrosian et al$.$ 1997). They have been commonly explained as
   infalling motions of HII regions \cite{saito}, multiple clouds
   merging \cite{skill93} and local peculiar gas motions induced by
   violent star formation events \cite{petro}. Very high star
   formation rates associated with these intense star forming events
   have been demonstrated to be able to produce a cavity of
   shock-heated gas due to the energy input provided by supernovae and
   stellar winds (Chevalier \& Clegg 1985; Vader 1986, 1987). This hot
   gas will accelerate the ambient interstellar medium resulting in a
   collective supernova-driven wind. In fact, several galactic
   supernova-driven winds phenomenae have been found to be associated
   with violent star formation places in dwarf galaxies (Roy et al$.$
   1991; Izotov et al$.$ 1996; Martin 1996, 1998, CM98
   hereafter). They have been generally detected as holes in the
   neutral hydrogen distribution (Puche et al$.$ 1992; Brinks 1994),
   bubbles or shells in H$\alpha$ emission (Marlowe et al$.$ 1995, MHW
   hereafter) or from their hot gas X-ray emission \cite{bomans}.
  
   The existence of these phenomenae could produce the loss of a
   significant fraction of the galaxy interstellar medium. Depending
   on the final destination of the accelerated gas, these structures
   could produce no mass loss, $blow$-$out$, only affecting the galaxy
   chemical evolution, or $blow$-$away$ processes, with a significant
   loss of interstellar mass (Young \& Gallagher 1990; CM98; Mac Low
   \& Ferrara 1998). Consequently, these supernova-driven galactic
   winds are accepted to be a key parameter in the dwarf galaxy
   formation (Silk et al$.$ 1987; Mori et al$.$ 1997) and evolution
   (MHW; Mac Low \& Ferrara 1998).

\begin{figure}
   \epsfxsize=80mm
   \epsfysize=80mm
   \epsffile{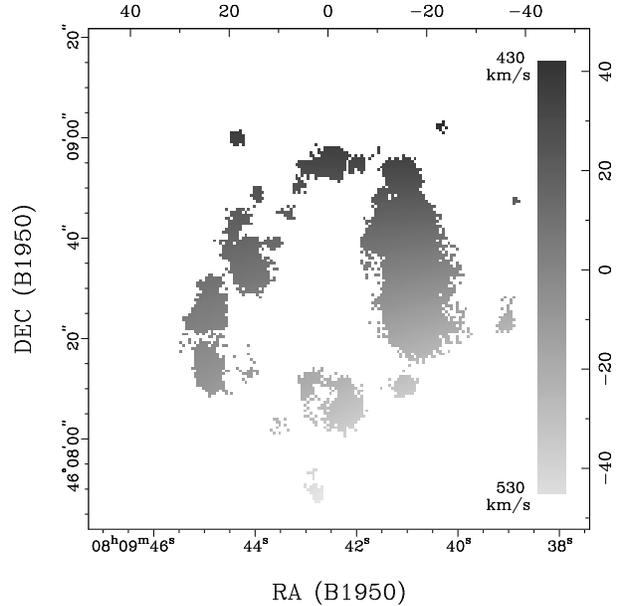}
\caption{Interpolated 2D velocity field. Only regions with $\Sigma_{\mathrm{H}\alpha}>$1.5$\times10^{-16}$\,erg\,s$^{-1}$\,cm$^{-2}$\,arcsec$^{-2}$ are shown.}

\label{field}
\end{figure}

   Blue Compact Dwarf galaxies, with intense recent or ongoing star
   forming activity, are those systems where the interplay between
   {\it star formation} and the {\it interstellar medium} is more
   feasible to be studied. However, although the majority of the BCD
   galaxies are iE type BCDs ($\simeq$70 per cent; Thuan 1991), with
   star formation spreading over the whole galaxy, the effects of the
   supernova-driven winds have been mainly studied in dwarf amorphous
   galaxies (see, e.g. MHW), which show nuclear star forming activity.

   The galaxy Mrk86=NGC~2537 (Shapley \& Ames 1932; Markarian 1969),
   also known as Arp 6 \cite{arp66}, constitutes an excellent
   laboratory to test the properties and effects of the
   supernova-driven winds on the interstellar medium of dwarf
   galaxies, as a nearby prototype of the iE BCD galaxies class.

\begin{figure*}
    \epsffile{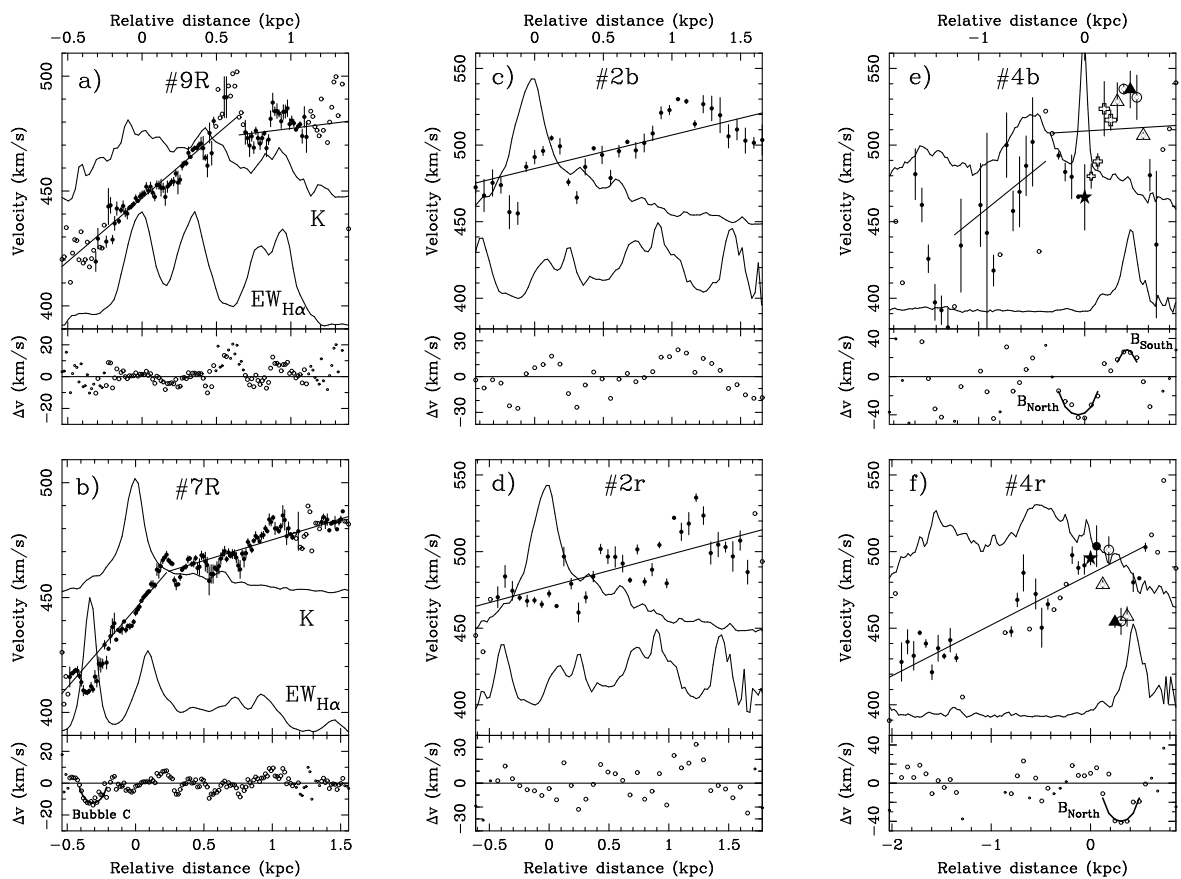} \epsfxsize=140mm
\caption{Heliocentric velocity profiles compared
with the $K$-band and EW$_{\mathrm{H}\alpha}$ profiles: {\bf a)} Slit
\#9R. {\bf b)} Slit \#7R. {\bf c)} Slit \#2b. {\bf d)} Slit \#2r. {\bf
e)} Slit \#4b. {\bf f)} Slit \#4r. Special symbols used in panels {\bf
e} and {\bf f} correspond to those regions marked in the left panel of
the Figure~\ref{bubbleB}. Open circles are velocity data obtained from
just one emission line, meanwhile filled circles are those measured
using several emission lines. The peak in the EW$_{\mathrm{H}\alpha}$
profile of the panel {\bf a)} has been taken as reference for relative
distances in this figure and it is due to the contribution from the
close knot \#16 (G99). The same for the panels {\bf b}, {\bf c} and
{\bf d} and the knot \#15 (see Figure~\ref{Aregion}). Distances in the
panels {\bf e} and {\bf f} are referred to the slit region closer to
the field star position (see Figure~\ref{bubbleB}). Lower panels show
the residuals from the global velocity gradient fitted.}

\label{profiles}
\end{figure*}

   After introducing Mrk86 in Sect.~\ref{mrk86}, we describe the
   observations and data reduction in
   Sect.~\ref{observations}. Results on the global velocity field of
   Mrk86 are given in Sect.~\ref{global}. In Sect.~\ref{models} we
   describe the evolutionary synthesis models applied. Then, we show
   the physical properties of the Mrk86--A (CM98) expanding bubble
   (Sect.~\ref{A}), and the new bubbles detected Mrk86--B and Mrk86--C
   (Sect.~\ref{B} \&~\ref{C}). The velocity dispersion measured in
   Mrk86--C is also discussed in Sect.~\ref{sigma}. Finally, summary
   and conclusions are given in Sect.~\ref{summary}. We have used
   $H_{\mathrm{0}}$=50\,km\,s$^{-1}$\,Mpc$^{-1}$ and
   $q_{\mathrm{0}}$=0.5 through this paper.

\section[]{Mrk86}
\label{mrk86}
   The galaxy Mrk86 ($\alpha$ (1950)= 08$^{h}$\,09$^{m}$\,43$^{s}$;
   $\delta$ (1950)= +46\degr\,08\arcmin\,33\arcsec), as a prototype of
   the iE type BCD galaxies, is characterized by a smooth elliptical
   low surface brightness underlying stellar component on which
   several irregular knots of star formation are superimposed. See
   Figure~\ref{image} for a $B$-band image. Up to 50 different knots
   have been detected (Gil de Paz et al$.$ 1999, G99 hereafter). In
   the Figure~\ref{field} we show those line-emitting regions with
   H$\alpha$ surface brightness higher than
   1.5$\times10^{-16}$\,erg\,s$^{-1}$\,cm$^{-2}$. Assuming a receding
   velocity of 443\,km\,s$^{-1}$ (de Vaucouleurs \& Pence 1980), the
   distance to Mrk86 would be 8.9\,Mpc and the physical scale
   42.9\,pc/\arcsec.  Mrk86 has an absolute magnitude of
   M$_{B}$=$-17^{\mathrm{m}}$, and it has been detected on 1.2, 2.8
   and 6.3 cm (Klein, Wielebinski \& Thuan 1984, Klein, Weiland \&
   Brinks 1991); the galaxy is a source of HI 21 cm emission (Thuan \&
   Martin 1981; Bottinelli et al$.$ 1984; WHISP survey, Kamphuis,
   Sijbring \& van Albada 1996); also it was detected in CO (Verter
   1985; Sage et al$.$ 1992); it is a strong IRAS source (Lonsdale et
   al$.$ 1985; Dultzin-Hacyan, Masegosa \& Moles 1990); near-infrared
   data were obtained by Thuan \shortcite{thu83}; optical images were
   obtained by Hodge \& Kennicutt \shortcite{hodge}, Loose \& Thuan
   \shortcite{loose} and Ojha \& Joshi \shortcite{oj:jos}; and,
   finally, ultraviolet spectra has been analyzed by Fanelli,
   O'Connell \& Thuan \shortcite{fan:con} and Longo et al$.$
   \shortcite{lon:cap}.

   Papaderos et al$.$ (1996a, 1996b; P96a \& P96b hereafter) obtained
   the physical parameters of the $plateau$, $central$ and
   $exponential$ photometric components. These components have been
   related with three different stellar populations (respectively,
   currently forming, $\simeq$1\,Gyr and $\geq$7\,Gyr old; Gil de Paz,
   Zamorano \& Gallego 1998, G98 hereafter).

\begin{table}
\caption[]{Journal of observations.}
\begin{tabular}{lrlcc}
&    \multicolumn{4}{c}{Spectroscopic observations}\\
          & Exp.  & Slit & Spectral & Disper. \\ 
Telescope & time (s) &  & range(nm) & (\AA /pix.) \\
\hline
CAHA 2.2-m & 3600 & 1,2,4,6b   & 330-580 & 2.65 \\
CAHA 2.2-m & 1800 & 5b         & 330-580 & 2.65 \\
CAHA 2.2-m & 3600 & 1,2,4,5,6r & 435-704 & 2.65 \\
CAHA 2.2-m & 3600 & 3          & 390-650 & 2.65 \\
INT 2.5-m  & 1800 & 7,8R       & 637-677 & 0.39 \\ 
INT 2.5-m  & 900  & 9R         & 637-677 & 0.39 \\ 
\hline
&    \multicolumn{4}{c}{Image observations}\\
          & Exp & Filter & Scale            & PSF  \\
Telescope & time (s) &        & (\arcsec /pixel) & (arcsec) \\
\hline
JKT 1.0-m & 600 & $B$ & 0\farcs33 & 1\farcs0 \\ 
CAHA 1.5-m & 2400 & $V$ & 0\farcs33 & 1\farcs6 \\ 
CAHA 2.2-m & 600 & Gunn-$r$ & 0\farcs27 & 1\farcs9 \\ 
INT 2.5-m & 900 & H$\alpha$ & 0\farcs57 & 2\farcs3 \\ 
INT 2.5-m & 900 & [OIII] & 0\farcs57 & 2\farcs5 \\ 
UKIRT 3.8-m & 100 & $K$ & 0\farcs60 & 1\farcs4 \\
\hline
\label{log}
\end{tabular}
\end{table}

\section{Observations and reductions}
\label{observations}
   We have obtained a total of 14 long-slit optical spectra in 10
   different orientations and positions (see
   Figure~\ref{image}). Journal of spectroscopic observations is given
   in Table~\ref{log}. Low-intermediate resolution spectra (FWHM=6\AA\
   in the light of H$\alpha$) with dispersion 2.65\,\AA/pixel were
   obtained with the Boller \& Chivens spectrograph at the Cassegrain
   focus of the 2.2-m telescope at Calar Alto (Almer\'{\i}a, Spain) in
   January 1993. We used a 1024$\times$1024 Tek 24$\mu$m CCD. High
   resolution spectra (FWHM=0.9\AA\ in H$\alpha$ and 0.39\,\AA/pixel;
   slits \#7R, \#8R and \#9R) were obtained using the IDS instrument
   at the Isaac Newton Telescope (INT) of the Observatorio del Roque
   de los Muchachos (La Palma, Spain) in January 1998. The detector
   was a 1024$\times$1024 Tek 24$\mu$m CCD. The spectra were reduced,
   wavelength and flux calibrated, using standard {\sc figaro}
   (January 1993) and IRAF\footnote{IRAF is distributed by the
   National Optical Astronomy Observatories, which are operated by the
   Association of Universities for Research in Astronomy, Inc., under
   cooperative agreement with the National Science Foundation.} 
   procedures (January 1998), and making use of the corresponding
   standard stars and lamp calibration spectra. Line fluxes were
   measured using the IRAF task {\sc splot}, obtaining errors lower
   than 15 per cent.

\begin{table}
\centering
\caption[]{Velocity gradients. $\omega_{\mathrm{0}}$ are the velocity gradients measured along the slits, meanwhile $\omega$ are the values projected to PA=12\degr. Distances along the rotation axis, d, have been measured relative to the $r$-band outer isophotes centre. The central value has been estimated from the interpolated 2D velocity map.}
\begin{tabular}{cccc}
\hline
 d (kpc) & $\omega_{\mathrm{0}}$ (km\,s$^{-1}$\,kpc$^{-1}$) & $\omega$
 (km\,s$^{-1}$\,kpc$^{-1}$) & Slit \\
         &   Steep/flat comp.  & Steep/flat comp. & \\
\hline
 1.03  & 55$\pm$4/8$\pm$3 & 56$\pm$5/8$\pm$3 & \#9R \\
 0.85  & 69$\pm$5/18$\pm$2 & 70$\pm$5/19$\pm$2 & \#7R \\
 0.73  & 21$\pm$4 & 21$\pm$4 & \#2r \\
 0.73  & 22$\pm$5 & 22$\pm$5 & \#2b \\
 0.00  & & 34$\pm$5 &  \\
$-$0.12  & 33$\pm$4 & 33$\pm$4 & \#4r \\
\hline
\label{gradients}
\end{tabular}
\end{table}

   We have also obtained $B$, $V$, $r$ and $K$-band images in
   different observing runs as described in Table~\ref{log}. Gunn-$r$
   band image was obtained at the 2.2-m telescope at Calar Alto in
   February 1992 with a GEC CCD with 800$\times$1156 30$\mu$m
   pixels. The Johnson-$V$ band image was obtained at the 1.52m
   telescope at Calar Alto in December 1993 with a 1024$\times$1024
   19$\mu$m pixels CCD. $B$ band image was obtained at the 1-m Jacobus
   Kapteyn Telescope (JKT) at the Observatorio del Roque de los
   Muchachos in November 1997 with a 24$\mu$m 1024$\times$1024 pixels
   Tek CCD. Finally, the near-infrared $K$-band image was obtained at
   the 3.8-m UKIRT telescope at Mauna Kea observatory in April 1993. We
   used the 62$\times$58 InSb array camera IRCAM. Broad band images
   were reduced using standard MIDAS and IRAF procedures. The
   procedures described by Arag\'on-Salamanca et al$.$
   \shortcite{alfonso} were employed for the reduction of the $K$-band
   near-infrared image. All the broad-band images were calibrated
   observing standard stars at different airmasses.

   Additional narrow-band images were secured for us during service time  
   at the 2.5-m Isaac Newton Telescope at the Observatorio del Roque de 
   los Muchachos in December 1993. The detector was an EEV CCD with 
   1280$\times$1180 22.5$\mu$m pixels. [OIII]5007\AA\ 
   ($\lambda_{\mathrm{0}}$=5014\AA; FWHM=50\AA) and on/off H$\alpha$ 
   ($\lambda_{\mathrm{0}}^{\mathrm{on}}$=6556\AA, FWHM=60\AA; 
   $\lambda_{\mathrm{0}}^{\mathrm{off}}$=6607\AA, FWHM=53\AA) images were 
   obtained. The continuum subtraction was performed using, respectively 
   for [OIII] and H$\alpha$, the $V$-band and off-H$\alpha$ images.

   The [OIII] and H$\alpha$ images were flux calibrated as
   follows. First, we convolved the $b$ ($blue$-$arm$; see
   Table~\ref{log}) spectra with the [OIII] filter and the $r$
   ($red$-$arm$) spectra with the H$\alpha$ filter (see
   Table~\ref{log} \& Figure~\ref{image}). Then, we collapsed these
   spectra in wavelength, obtaining the flux calibrated spatial
   profile of the [OIII] and H$\alpha$ emission. Now, we cropped and
   averaged in the [OIII] and H$\alpha$ images those regions covered
   by the $b$ and $r$ slits, obtaining the [OIII] and H$\alpha$
   spatial profiles given by the images. Comparing these spatial
   profiles with those obtained from the spectra we calibrated both
   the [OIII] and H$\alpha$ images. Finally, these calibration
   relations were corrected for the sensitivity of the filters at the
   corresponding wavelength and, in the case of the H$\alpha$
   emission, from the contribution of the [NII]6548\AA\, and 6583\AA\,
   lines. The reliability of this method was demonstrated after
   applying it to all the $blue$-$arm$ and $red$-$arm$ spectra,
   obtaining similar results within an error of 10 per cent.

\section{Global velocity field}
\label{global}

\begin{figure}
   \epsfxsize=80mm
   \epsfysize=80mm
   \epsffile{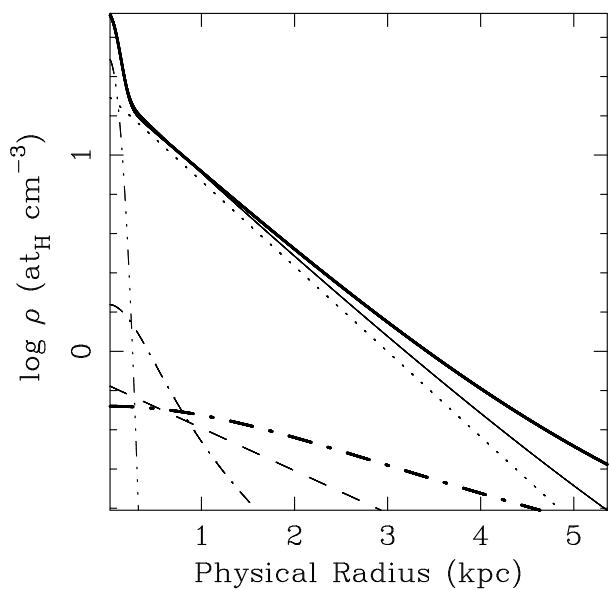}
\caption{{\bf Thin lines}: Neutral hydrogen ($dashed$), stellar-underlying 
($dotted$), stellar-starburst ($dot$-$dot$-$dot$-$dashed$), dark
matter ($dot$-$dashed$) and total ($solid$) mass density profiles for
$R_{\mathrm{DM}}$=0.5\,kpc. {\bf Thick lines}: Dark matter
($dot$-$dashed$) and total ($solid$) mass density profiles for
$R_{\mathrm{DM}}$=3.0\,kpc.}

\label{rho}
\end{figure}

   We have obtained the spatial variation of the ionized gas
   velocities using the long-slit spectra. We have used the {\sc
   rvidlines} IRAF task measuring the [OII]3727\AA, H$\beta$,
   [OIII]4959\AA\, and [OIII]5007\AA\, line velocities for the
   $blue$-$arm$ spectra, and [OIII]4959\AA, [OIII]5007\AA,
   [NII]6548\AA, H$\alpha$, [NII]6583\AA, [SII]6717\AA\, and
   [SII]6731\AA\ for the $red$-$arm$ spectra, weighted with their
   relative intensities. No significant differences were observed
   using allowed and forbidden lines separately. Errors in the
   velocity were estimated there where several emission lines could be
   measured (see Figure~\ref{profiles}).
  
   From the velocities determined in the spectra, and the slit
   positions given in Figure~\ref{image}, we have reconstructed a 2D
   velocity map (see Figure~\ref{field}) using the IRAF task {\sc
   xyztoim}. A careful examination of the interpolated 2D velocity
   field yield a global {\it solid-body} velocity field of maximum
   velocity gradient $\Omega$=34\,km\,s$^{-1}$\,kpc$^{-1}$ and
   orientation PA$\simeq$12\degr\ (rotation axis
   PA$\simeq-$78\degr). The heliocentric velocity measured at the
   galactic centre, as given by the $r$-band outer isophotes centre,
   is about 470\,km\,s$^{-1}$. Although De Vaoucoleurs \& Pence
   \shortcite{vauco} gave a median heliocentric velocity of
   443\,km\,s$^{-1}$, a detailed examination of their Fabry-Perot
   interferogram shows that the velocity close to the galactic center
   is significantly higher, between 450-460\,km\,s$^{-1}$.

   In Figure~\ref{profiles} we show the velocity profiles measured
   along the slits \#9R, \#7R, \#2b, \#2r, \#4b and \#4r, those
   positioned closer to PA=12\degr. The six panels have been arranged
   in slit position order, from West to East, where slit \#4b is that
   closely crosses the galactic centre, as defined by the $r$-band
   outer isophotes ($\alpha$ (1950)= 08$^{h}$\,09$^{m}$\,42\fs56;
   $\delta$ (1950)= +46\degr\,08\arcmin\,33\farcs8). Northernmost part
   of the slit is represented at the leftside of each panel.

   The panels {\bf a)} and {\bf b)} (slits \#9R and \#7R) show two
   different velocity components, one very steep gradient and a second
   flatter velocity gradient. The steeper components seem to be
   associated with regions of enhanced $K$-band luminosity and
   relatively high star forming activity 
   (EW$_{\mathrm{H}\alpha}\!\sim$100\AA), the emission knots \#15
   ($\alpha$ (1950)= 08$^{h}$\,09$^{m}$\,41\fs17; $\delta$ (1950)=
   +46\degr\,08\arcmin\,45\farcs1; G99) and \#16 ($\alpha$ (1950)=
   08$^{h}$\,09$^{m}$\,41\fs00; $\delta$ (1950)=
   +46\degr\,08\arcmin\,43\farcs0; G99; see
   Figure~\ref{profiles}). This increase in the radial velocity
   gradient could be due to an enhancement of the mass
   density over the global mass density distribution, related with the
   presence of these massive star forming regions. However, a merging
   with another dwarf galaxy or gas cloud with an independent velocity
   field, as it has been proposed to explain the velocity field of
   I~Zw~18 \cite{skill93} and II~Zw~40 \cite{vanzee}, should not be
   ruled out. In that case, this merging could be responsible for the
   triggering of the star formation in the emission knots \#15 and
   \#16 (G99).

\begin{figure}
\psfig{file=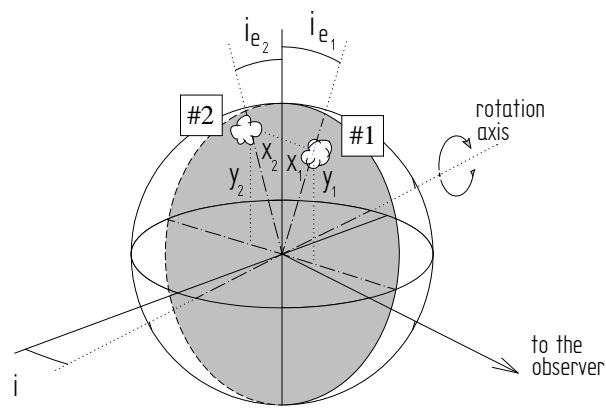,width=8.8cm,clip=}
\caption{Position angles, $i$ and $i_{\mathrm{e}}$, and coordinates $x$ and $y$ for two equatorial HII regions (\#1 \& \#2). Solid plane represents the galactic equator.}

\label{picture}
\end{figure}

   On the other hand, the \#9R and \#7R flat components and the
   gradients measured along the slits \#2b, \#2r and \#4r should be
   related with a more relaxed global velocity field. In
   Table~\ref{gradients} we give these velocity gradients, being
   $\omega_{\mathrm{0}}$ the gradients measured along the slits and
   $\omega$ the values projected to PA=12\degr, i.e.
   $\omega_{\mathrm{0}}$=$\omega\times\cos$(PA$-$12\degr). Velocities
   affected by local motions, such as supernova-driven winds (see
   Sect.~\ref{evidences}) have not been included in the calculus of
   the velocity gradients. From Table~\ref{gradients} we see that the
   flat component of the velocity gradient grows from
   10-20\,km\,s$^{-1}$\,kpc$^{-1}$ in the outer galaxy regions to
   $\sim$34\,km\,s$^{-1}$\,kpc$^{-1}$ close to the galactic centre.

\subsection{Mass density profile}
\label{dprofile}
   We will assume that the motion of the HII regions which produce the
   line-emission observed is due to rotation. The slit \#4r was placed
   very close ($\sim$3\arcsec) along the galactic equator. Therefore,
   we could compare its velocity profile with the radial component of
   the circular velocity curve \cite{binney}. We will parameterize the
   mass distribution of the galaxy (stellar and dark matter, and
   neutral and molecular hydrogen). We have assumed that the
   photometric centre, given by the $r$-band outer isophotes,
   coincides with the kinematical centre (see \"{O}stlin et al$.$
   1998). In order to compare the velocity curve obtained along the
   slit \#4r with the radial component of the circular velocity curve
   we will adopt the radial velocity measured close to the galactic
   center as the sistemic velocity.\\

\noindent {\bf Stellar}. From the analysis of the surface 
   brightness profiles of 14 Blue Compact Dwarf galaxies, P96a found
   that these profiles can be fitted using three distinct components,
   the {\it underlying}, {\it plateau} and {\it starburst}
   components. Using a numerical deprojection procedure, they obtained
   the central luminosity densities and scale lengths for these
   components. G98 estimated the ages for the {\it starburst} and {\it
   underlying} components to be, respectively, 1 and 7\,Gyr, with one
   fifth solar metallicity. Then, we have taken the mass-to-light
   ratios predicted by the Bruzual \& Charlot \shortcite{bc96}
   evolutionary synthesis models for these ages and metallicities,
   under the assumption of instantaneous star formation and using an
   Scalo IMF \cite{scalo}. These mass-to-light ratios result in
   M/L$_B$=0.5\,M$_{\sun}$/L$_{\sun}$ for the {\it starburst}
   component and M/L$_B$=2.6\,M$_{\sun}$/L$_{\sun}$ for the {\it
   underlying} component. Although quite significant in luminosity,
   the {\it plateau} component represents a very small fraction in
   mass because its luminosity is produced by regions with star
   formation more recent than 20\,Myr (G99) and mass-to-light ratios
   lower than M/L$_B$=0.08\,M$_{\sun}$/L$_{\sun}$ \cite{bc96}.

   Finally, using the parameters given by P96a for the Mrk86 {\it
   starburst} and {\it underlying} components we obtain
\begin{eqnarray}
\rho^{under}(r) = \rho^{under}_0 \exp{(-r/R_{under})} \\
\rho^{sb}(r) = \rho^{sb}_0 \exp{(-r/R_{sb})^2} 
\end{eqnarray}
where\\ 
$\rho^{under}_0$=20\,at$_{\mathrm{H}}$\,cm$^{-3}$\\ 
$R_{under}$=1.0\,kpc\\ 
$\rho^{sb}_0$=31\,at$_{\mathrm{H}}$\,cm$^{-3}$\\ 
$R_{sb}$=0.14\,kpc\\
\vspace{0.cm} 

\begin{figure}
   \epsfxsize=80mm \epsfysize=80mm \epsffile{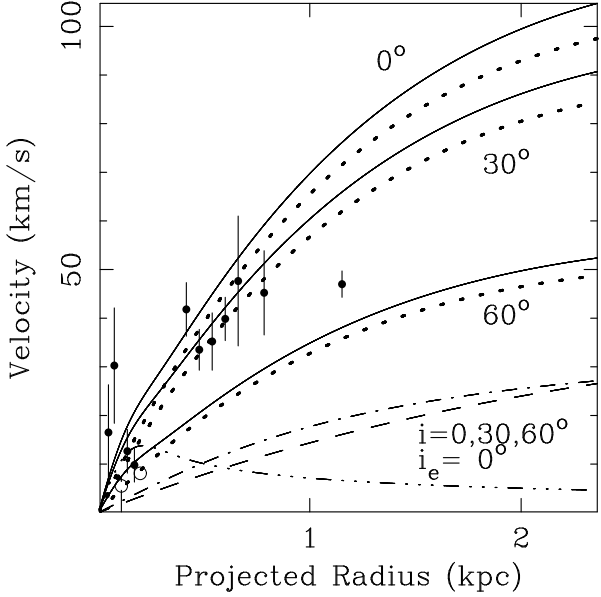}
\caption{Total ($solid$) and stellar ($dotted$) radial velocity curve predicted for $i_{\mathrm{e}}$=0\degr\ and inclination of the rotations axis, $i$, 0, 30 and 60\degr. Black and white dots are, respectively, receding and approaching velocities. The $dot$-$dot$-$dot$-$dashed$, $dot$-$dashed$ and $dashed$ lines, respectively the starburst, DM and HI density components are represented just for $i$=0\degr\ and $R_{DM}$=1\,kpc.}

\label{rvcurve1}
\end{figure}

\begin{figure}
   \epsfxsize=80mm \epsfysize=80mm \epsffile{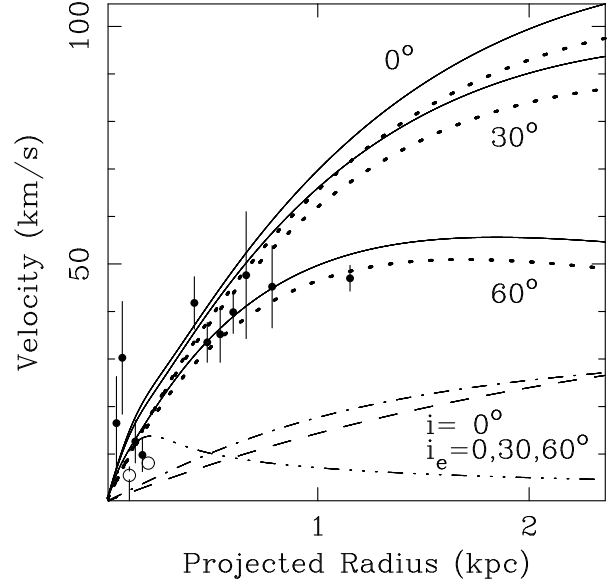}
\caption{Radial velocity curves predicted for $i$=0\degr\ and $i_{\mathrm{e}}$=0,30,60\degr. The meaning of the symbols and line patterns are the same that in Figure~\ref{rvcurve1}. Starburst, DM and HI density components are represented just for $i_{\mathrm{e}}$=0\degr\ and $R_{DM}$=1\,kpc.}

\label{rvcurve2}
\end{figure}

\noindent   {\bf Neutral hydrogen}. If we assume that the HI distribution, 
   as given by the WHISP survey\footnote{\tt
   http://thales.astro.rug.nl/$\sim$whisp/Database/}, falls
   approximately as an exponential function, and using the same
   deprojection procedure that for the optical data (see P96a), we
   obtain
\begin{equation}
\rho^{\mathrm{HI}}(r) = \rho^{\mathrm{HI}}_0 \exp{(-r/R_{\mathrm{HI}})}
\label{rhoHI} 
\end{equation}
where\\
$\rho^{\mathrm{HI}}_0\!\simeq$0.7\,at$_{\mathrm{H}}$\,cm$^{-3}$\\
$R_{\mathrm{HI}}\!\simeq$2.0\,kpc\\

\noindent   {\bf Molecular hydrogen}. Sage et al$.$ \shortcite{sage} gave a 
   M(H$_{2}$)/M(HI) ratio for 
   Mrk86 of 0.03. Therefore, we can assume negligible the effect of 
   the molecular hydrogen in the gravitational potential of Mrk86.\\

\noindent   {\bf Dark matter}. Finally, we have considered the effect 
   of the dark matter distribution over the global velocity
   field. Following the most recent observational works (Salucci \&
   Persic 1997; Flores \& Primack 1994; Moore 1994) and the study of
   Navarro et al$.$ \shortcite{nef} concerning the effects of
   supernova-driven winds over standard CDM profiles, we have assumed
   the existence of a central {\it core} in our dark matter density
   profile. Both the density distribution of a modified isothermal
   sphere (see Binney \& Tremaine 1987) and the universal profile
   given by Burkert \shortcite{burkert} include this central {\it
   core} and reproduce reasonably the observations. We have used for
   this work the simplier modified isothermal profile, also following
   Mac Low \& Ferrara \shortcite{mclow},
\begin{equation}
\rho^{\mathrm{DM}}(r) = \frac{\rho^{\mathrm{DM}}_0}{1+ (r/R_{\mathrm{DM}})^2}
\label{rhoDM}
\end{equation}
    Burkert (1995; see also Mac Low \& Ferrara 1998) has shown that
    the central density $\rho^{\mathrm{DM}}_0$ is related with the
    scale radius $R_{\mathrm{DM}}$ through the expression
\begin{equation}
\rho^{\mathrm{DM}}_0 = 2.7\times10^7\,\left( {R_{\mathrm{DM}} \over \mathrm{kpc}} \right) ^{-2/3} \ \  M_{\sun}\,\mathrm{kpc}^{-3}
\label{burkerteqn}
\end{equation}
   Thus, this profile has only one free parameter, $R_{\mathrm{DM}}$.

   In Figure~\ref{rho} we show the total and DM mass density profiles 
   for $R_{\mathrm{DM}}$=0.5\,kpc and 3\,kpc. In both cases we see
   that the total mass density profile is dominated by the stellar
   component within the central 4\,kpc.

\begin{figure}
\psfig{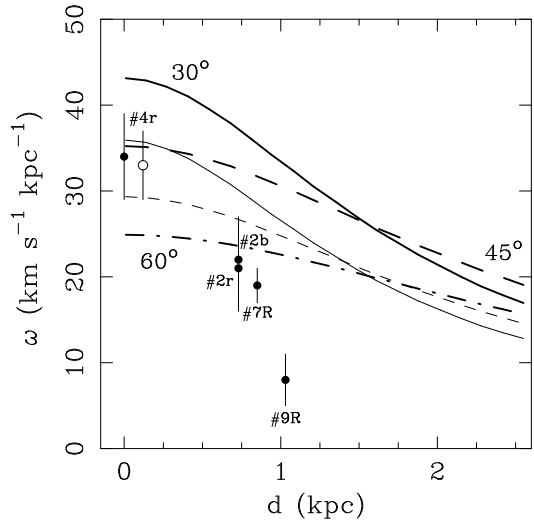}
\caption{Change in the radial velocity gradients with the distance along the rotation axis. Data points have been taken from Table~\ref{gradients}. For d=0 we have taken the velocity gradient measured from the interpolated 2D velocity map. $Thick$-$lines$ represent the velocity gradient predictions for 30\degr\ ($solid$), 45\degr\ ($dashed$) and 60\degr\ ($dot$-$dashed$) inclined rotation axis with the density profile given in Sect.~\ref{dprofile}. $Thin$-$lines$ are the predictions for $i$=30\degr\ ($solid$) and 45\degr\ ($dashed$), but using $R_{under}$=0.5\,kpc, instead of 1.0\,kpc. White dot was obtained for a projected distance of $-$0.12\,kpc east of the galactic centre (see Figure~\ref{image}).}

\label{gradgrad}
\end{figure}

   Now, we could compare the radial component of the circular velocity
   curve derived from the total density profile with the radial
   velocity data obtained along the slit \#4r, placed close along the
   galactic equator. However, since the ionized gas emission is
   produced in several individual HII regions, the projected radial
   component of the modelized circular velocity will depend on the
   position of the corresponding HII region.

\subsection{Ionized gas geometry}
\label{distribution}

   We will compare the velocity data obtained along the equatorial
   slit \#4r with the projection of the rotation curve adopting
   several geometries. 

   We will assume that the ionized gas is distributed in a {\it
   thin-disk} with inclination $i$. This angle measures the
   inclination of its rotation axis with regard to the plane of the
   sky (i.e. $i$=0\degr\ for a edge-on disk; see
   Figure~\ref{picture}). In the Figure~\ref{picture} we also show how
   the angles $i$ and $i_{\mathrm{e}}$ and the coordinates $x$ and $y$
   describe the position in the disk of an HII region. Under the
   hypothesis of {\it thin-disk}, if $i\!\neq$0\degr\ only those regions
   with angle $i_{\mathrm{e}}\!\simeq$0\degr\ will be observed through
   the slit \#4r. This is the case shown in the Figure~\ref{rvcurve1}.

   However, if the galaxy would have an inclination close to
   $i$=0\degr, HII regions with different position angle,
   $i_{\mathrm{e}}$, could be observed through the slit. In this
   situation, we could observe regions with different radial velocity
   at the same apparent position (regions \#1 \& \#2 in
   Figure~\ref{picture}). For example, if we consider the mass density
   profile given in Sect.~\ref{dprofile} (with
   $R_{\mathrm{DM}}$=1\,kpc), we could measure changes in the radial
   velocity of 15\,km\,s$^{-1}$ between two regions with $x$=0\,kpc
   and $x$=3\,kpc for $y$=2.5\,kpc (see
   Figure~\ref{picture}). Therefore, the existence of inhomogeneities
   in the ionized gas distribution in the galaxy could also produce
   short scale variations in its velocity field.

   The emitting HII regions can have, in principle, any position angle
   $|i_{\mathrm{e}}|\leq$180\degr. In fact, a region with a given
   $i_{\mathrm{e}}$ at present, would evolve toward larger
   $i_{\mathrm{e}}$ angles in the future. Had all the HII regions be
   placed in planes of constant angle $i_{\mathrm{e}}$, the resulting
   rotation curves would be the ones shown in Figure~\ref{rvcurve2}.

   From the Figure~\ref{rho} we see that the stellar mass density
   component dominates at the inner 4\,kpc. However, the radial
   velocity data can be fitted adopting different geometries for the
   ionized gas distribution (see Figures~\ref{rvcurve1}
   and~\ref{rvcurve2}). We could reproduce the velocities measured if
   $i$=0\degr\ and 30\degr$<i_{\mathrm{e}}<$70\degr, or if
   $i\!\simeq$40\degr and $i_{\mathrm{e}}$=0\degr.

   On the other hand, the way the velocity gradient perpendicular to
   the rotation axis, $\omega$, decrease from the equator to galactic
   outer regions (see Table~\ref{gradients}) resembles the velocity
   fields observed in spiral galaxies with intermediate inclination
   (e.g. Giovanelli \& Haynes 1988). We will test this point assuming
   that the emission observed effectively comes from a rotating {\it
   thin-disk} with gravitational potential that given by the mass
   distribution described before (Sect.~\ref{dprofile}). Then, using
   this gravitational potential we have estimated the expected radial
   velocity gradients for different distances, d, along the rotation
   axis (see Table~\ref{gradients}). In Figure~\ref{gradgrad} we show
   the gradients expected, measured as the change in radial velocity of
   their inner two kiloparsecs. Data points have been taken from
   Table~\ref{gradients}. The value obtained for the slit \#9R is very
   uncertain since the small number of points used to fit its velocity
   gradient (see Figure~\ref{profiles}). The zero value represents the
   gradient along the galactic equator obtained from the interpolated
   2D velocity map. From this figure we note that, although a
   decreasing in the velocity gradient is effectively observed, we
   cannot reproduce the values measured. The more feasible explanation
   is that the ionized gas is probably distributed in a relatively
   {\it thick-disk}. However, the use of a steeper mass density
   profile could also reproduce the velocity gradients measured (see
   Figure~\ref{gradgrad}).

   Therefore, we conclude that the observed global velocity field of
   Mrk86 can be reproduced if the stellar component dominates the
   total mass profile within its optical radius, and if the emitting
   ionized gas is distributed in a probably thick inclined disk. The
   inclination of the disk with regard to the plane of the sky will be
   about 50\degr\ ($i\!\simeq$40\degr).

\section{Kinematical evidences for supernova-driven winds}
\label{evidences}

   Several under-kiloparsec-scale kinematical structures are observed
   superimposed on the global velocity gradients previously
   described. These kind of kinematical disturbances have been widely
   observed in dwarf galaxies (see Tomita et al$.$ 1997; Petrosian et
   al$.$ 1997). Different mechanisms could properly explain these
   disturbances, including infalling motions of HII regions
   \cite{saito} or inhomogeneities in the ionized gas distribution
   (see Sect.~\ref{global}). However, in our case, the deep minima
   (see
   Figures~\ref{profiles}b,~\ref{profiles}e,~\ref{profiles}f,~\ref{Cratios}a
   and~\ref{Cratios}b) and velocity maximum (Figure~\ref{profiles}e)
   observed are clearly spatially correlated with intense star forming
   knots, indicating the existence of supernova-driven winds (see
   Sect.~\ref{A},~\ref{B} \&~\ref{C}).

\subsection{Description of the evolutionary synthesis models}
\label{models}
   In order to determine the physical properties of the star forming
   knots, in particular, those associated with supernova-driven wind
   phenomenae, we have made use of the Leitherer and Heckman (1995; LH
   hereafter) models. These single burst models do not take into
   account any contribution from the old underlying population neither
   that from emission lines (i.e. [OIII]5007\AA, H$\alpha$).

   Optical and near infrared colours measured in the outer galaxy
   regions ($B-V$=0.8, \-$V-r$=\-0.0, $V-K$=2.5; G98), where the
   exponential underlying population dominates (P96a), allow to
   estimate an age of about 7\,Gyr (G98 using the Bruzual \& Charlot
   1996 models) for this underlying component. Assuming different
   burst-strength values, ranging from 0.001 (0.1 per cent in mass) to
   1, we have rebuilt a complete set of LH instantaneous burst models
   of 0.25 and 0.1\,Z$_{\sun}$ metallicities, Salpeter IMF
   \cite{salpeter} with m=$-$2.35, M$_{low}$=1\,M$_{\sun}$ and
   M$_{up}$=100\,M$_{\sun}$. Then, the relation \cite{kent},
\begin{equation} 
                    r = R + 0.41 + 0.21 \times (V - R) 
\label{gunnr}
\end{equation} 
   has been applied in order to transform from Johnson-$R$ to Gunn-$r$
   magnitudes. Finally, using the H$\alpha$ equivalent widths measured
   from the spectra, we have corrected the Gunn-$r$ magnitudes from
   H$\alpha$ emission.

\subsection{Previously detected Mrk86--A bubble}
\label{A}
\begin{figure}
\psfig{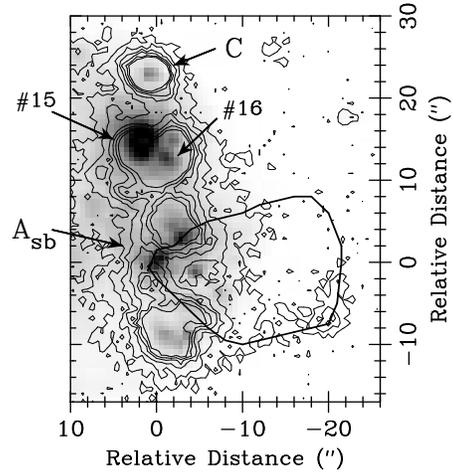}
\caption{Mrk86--A region. $B$ image and H$\alpha$
contours are shown. Displacements between $B$-band knots and H$\alpha$
maxima are evident in the position of the Mrk86--A bubble starburst
precursor (CM98), A$_{\mathrm{sb}}$. The limits of the Mrk86--A
H$\alpha$ emission given by CM98 are also shown, joint with the
positions of the \#15 and \#16 (G99) emission knots.}

\label{Aregion}
\end{figure}

   Using long--slit echelle spectra in H$\alpha$+[NII] with
   11\,km\,s$^{-1}$ resolution (FWHM), CM98 inferred the presence of
   an expanding bubble extending 22\arcsec\,(940\,pc) westward of the
   galaxy west end. The physical parameters of this bubble are given
   in Table~\ref{parameters}. In Figure~\ref{Aregion} we show the
   A-big bubble region in H$\alpha$ and $B$-band. Following CM98, we
   have taken the knot marked in Figure~\ref{Aregion} as
   A$_{\mathrm{sb}}$ (\#31 emission knot in G99) as the starburst
   precursor.

   For this region and for the Mrk86--B and C bubble precursors, we
   have determined physical apertures using the program {\sc cobra}
   (G99). This program determines and subtracts the underlying
   emission. The apertures include those pixels with emission more
   intense than 1/e times the emission maximum. Optical-nIR colours
   have been obtained for these apertures and corrected for internal
   extinction applying the relation given by Calzetti
   \shortcite{calze}, $\mathrm{E}(B-V)_{\mathrm{continuum}}$=0.44
   $\times \mathrm{E}(B-V)_{\mathrm{gas}}$. Using the
   H$\alpha$-H$\beta$ Balmer decrements and assuming a Mathis
   \shortcite{mathis} extinction law, colour excesses
   $\mathrm{E}(B-V)_{\mathrm{gas}}$ have been obtained. The corrected
   colours are given in Table~\ref{parameters}. These colours allow to
   estimate the {\it mean} physical properties, i.e. age, metallicity
   and burst strength for these regions. In Table~\ref{parameters} we
   show the physical parameters of the LH model that better fits the
   colours measured.

   In the Mrk86--A case, the colours measured yield a LH model with
   0.25\,Z$_{\sun}$ metallicity, burst strenght of 1 per cent and age
   of about 14\,Myr (see Table~\ref{parameters}). This value is
   slightly different from the dynamical time deduced by CM98
   (converted to $H_{\mathrm{0}}$=50\,km\,s$^{-1}$\,Mpc$^{-1}$), that
   is $t_{\mathrm{dyn}}\!\simeq$12\,Myr. This discrepancy could be
   explained taking into account that $t_{\mathrm{dyn}}$ describes the
   age of the expanding bubble, that differs from the starburst age in
   a time similar to the more massive stars main sequence phase
   duration. After the red super giant phase has started
   ($\sim$4$\times10^{6}$\,yr after the starburst for
   $M\!\simeq$40\,M$_{\sun}$; Maeder 1990), massive stars winds and
   later supernova explosions would commence to take place and a
   cavity of shock-heated gas could begin to form.

\subsection{Mrk86--B}
\label{B}
\begin{figure*}
    \epsfxsize=125mm
    \epsffile{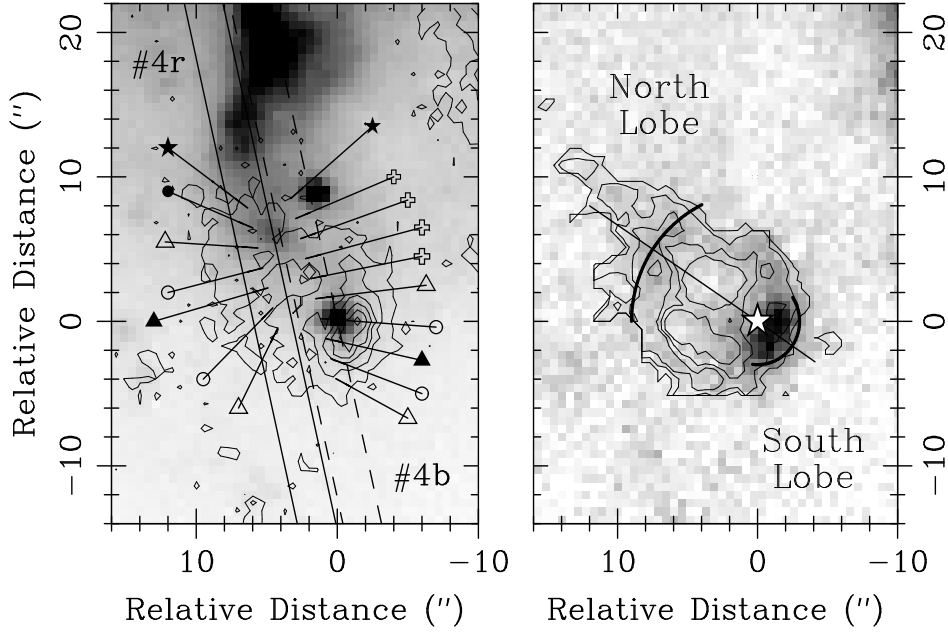}
\caption{{\it Left panel:} Mrk86--B region in B-band (image) and 
     H$\alpha$ (contours). Region 1\arcsec\ East and 9\arcsec\ North
     is a field star. {\it Right panel:} H$\alpha$ (image) and
     [OIII]/H$\alpha$ (contours). See Figures~\ref{profiles}e
     and~\ref{profiles}f for the velocities measured in the regions
     marked on the left panel. In the [OIII]/H$\alpha$ case, some
     contour-map highly noisy regions have been artificially removed.}

\label{bubbleB}
\end{figure*}

   A local velocity maximum and minimum is observed along the \#4b
   slit velocity profile (see Figure~\ref{profiles}e). They differ
   68\,km\,s$^{-1}$ in velocity and are 10\arcsec\,(430\,pc)
   apart. These features are compatible with the approaching (north
   lobe) and receding parts (south lobe) of a bipolar expanding
   structure with a characteristic projected expansion velocity of
   34\,km\,s$^{-1}$.  The approaching velocity minimum is also
   observed in the \#4r slit velocity profile (see
   Figure~\ref{profiles}f).  Associated with these kinematical
   features a clear bubble-like structure is observed in the H$\alpha$
   narrow band image (see Figure~\ref{bubbleB}). Assuming a distance
   of 8.9\,Mpc, a physical size of 750$\times$510\,pc$^{2}$ is
   obtained. This structure resembles those observed in I~Zw~18
   \cite{cm96}, M82 \cite{heck90} and NGC~1705 (MHW). The difference
   in size between both lobes could be related with a strong ambient
   density gradient, with higher density toward the south lobe region.

\begin{table*}
\begin{minipage}{150mm}
\centering
\caption[]{Bubbles physical parameters. dE/dt$_{\mathrm{kin}}$ values
    given in the {\it upper panel} are deduced applying the
    expressions of $r_{\mathrm{bubble}}$ and $v_{\mathrm{exp}}$ from
    Castor, McCray \& Weaver \shortcite{castor}. {\it Lower panel}
    shows the parameters of the best fitted evolutionary LH model and
    the predicted collisionally excited H$\alpha$
    luminosity. Optical-nIR colours in parenthesis ({\it upper panel})
    are those given by the corresponding LH
    model. E$(B-V)_{\mathrm{continuum}}$ is assumed to be
    0.44$\times$E$(B-V)_{\mathrm{gas}}$
    \cite{calze}. E$(B-V)_{\mathrm{gas}}$ values for Mrk86--A bubble
    have been obtained by CM98 and this work (in
    parenthesis). Following the Castor et al$.$ \shortcite{castor}
    bubble model, dynamical ages, $t_{\mathrm{dyn}}$, have been
    estimated as 0.6$\times$$r_{\mathrm{bubble}}$/$v_{\mathrm{exp}}$
    (see, e.g. CM98).}

\begin{tabular}{llcrcrcr}
 & & A(big)\footnote{CM98, converted to $H_{\mathrm{0}}$=50\,km\,s$^{-1}$\,Mpc$^{-1}$ } && B (N/S lobes) & & C \\
\hline
RA(B1950) && 08$^{h}$09$^{m}$40\fs0 && 08$^{h}$09$^{m}$42\fs3 && 08$^{h}$09$^{m}$41\fs1 & \\ 
DEC(B1950) && 46\degr 08\arcmin 25\farcs0 && 46\degr 08\arcmin 6\farcs5 && 46\degr 08\arcmin 52\farcs6 & \\ 
$r_{\mathrm{bubble}}$ & (pc) & 944 && 558/190 && 110 & \\ 
$v_{\mathrm{exp}}$ & (km\,s$^{-1}$) & 47 && 40/28 && 17 & \\ 
$t_{\mathrm{dyn}}$ & (Myr) & 12 && 6.5 && 4 &\\
dE/dt$_{\mathrm{kin}}$ & (erg\,s$^{-1}$) & 1.2$\times10^{40}$ && 2.5/0.1$\times10^{39}$ && 8$\times10^{36}$ & \\ 
L$_{\mathrm{H}\alpha}$ & (erg\,s$^{-1}$) & 7$\times10^{37}$ && 0.5/2.5$\times10^{38}$ && 1.7$\times10^{39}$ & \\
E$(B-V)_{\mathrm{gas}}$ & & 0.30 & (0.13) & 0.16 && 0.73 &\\ 
$m_{\mathrm{B}}$ (knot) & & 17.12 && 17.77 && 17.65 & \\ 
$B-V$ & & 0.22$\pm$0.06 & (0.27) & 0.17$\pm$0.08 & (0.19) & 0.13$\pm$0.08 & (0.12) \\ 
$V-r$ & & $-$0.04$\pm$0.02 & ($-$0.10) & $-$0.11$\pm$0.04& ($-$0.17) & 0.05$\pm$0.04 & (0.00) \\
$r-K$ & & 2.34$\pm$0.09 & (2.29) & 2.1$\pm$0.2 & (2.11) & 1.73$\pm$0.17 & (1.75) \\ 
\hline
Burst age & (Myr) & 14.2 && 11.5 && 7.7 &\\ 
Burst strength & & $\sim$0.01 && $\sim$0.01 && $\sim$0.01 &\\ 
Burst metallicity & (Z$_{\sun}$) & $\sim$0.25 && $\sim$0.10 && $\sim$0.10 & \\ 
dE/dt$_{\mathrm{kin}}$ & (erg\,s$^{-1}$) & 2.2$\times10^{39}$ && 1.4$\times10^{39}$ && 9$\times10^{38}$ \\
L$_{\mathrm{H}\alpha}$ coliss. & (erg\,s$^{-1}$) & 9$\times10^{36}$ && 2.4/1.7$\times10^{36}$ && 2.3$\times10^{35}$ & \\ 
Mass & (M$_{\sun}$) & 11$\times10^{4}$ && 6$\times10^{4}$ && 5$\times10^{4}$ &\\
\hline
\label{parameters}
\end{tabular}
\end{minipage}
\end{table*}
\begin{figure*}
\vspace{10cm}
\includegraphics{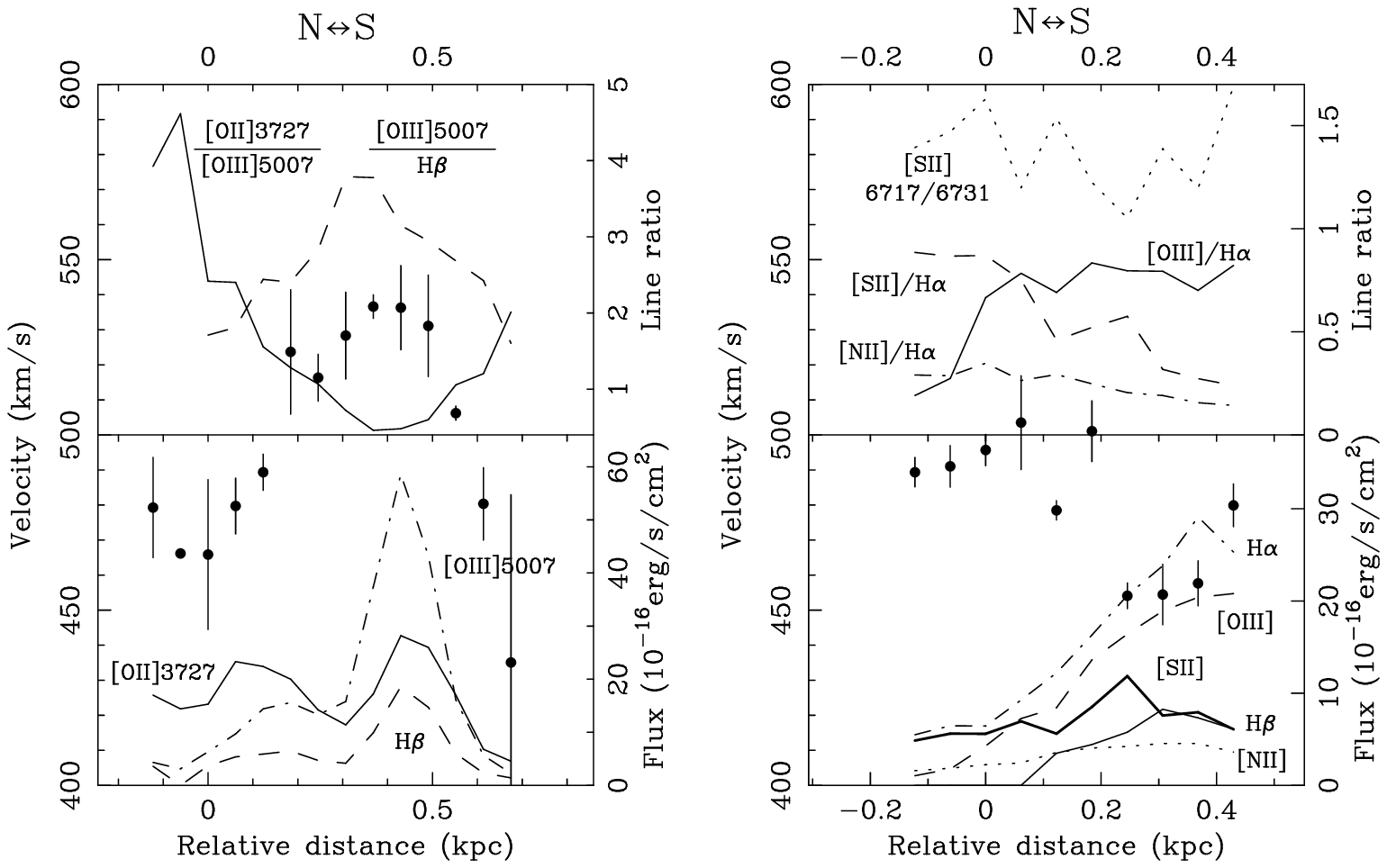}
\caption{Mrk86--B region. {\bf a)} {\it Filled dots:} \#4b 
long-slit velocity profile. {\it Upper panel:} Observed [OII]/[OIII]
profile ($solid$ line), [OIII]/H$\beta$ profile ($dashed$ line). {\it
Lower panel:} [OII] ($solid$), H$\beta$ ($dashed$) and [OIII]
($dot$-$dashed$) profiles. {\bf b)} {\it Filled dots:} \#4r long-slit
velocity profile. {\it Upper panel:} Observed [OIII]/H$\alpha$ profile
($solid$), [SII]/H$\alpha$ profile ($dashed$), [NII]/H$\alpha$ profile
($dot$-$dashed$) [SII]6717\AA/[SII]6731\AA\, profile ($dotted$). {\it
Lower panel:} H$\beta$ ($solid$), [OIII] ($dashed$), H$\alpha$
($dot$-$dashed$), [NII] ($dotted$), [SII] ($thick$ $solid$). Relative
distances are referred to the same regions as in
Figures~\ref{profiles}e
\&~\ref{profiles}f.}

\label{Bratios}
\end{figure*}

   At 3\arcsec\,(130\,pc) to the NE from the bubble geometrical centre
   (origin of coordinates in Figure~\ref{bubbleB}) an optical emission
   knot (\#49 in G99) is observed in the $B$, $V$, $r$ and $K$ bands
   (Figure~\ref{bubbleB}, left panel). If we compare the instantaneous
   burst evolutionary synthesis LH models to the colours measured (see
   Sect.~\ref{models} and Table~\ref{parameters}), considering an age
   for the old underlying population of 7\,Gyr (G98), the best fit
   obtained yields an starburst age of about 11.5\,Myr, a burst
   strength of 1 per cent and a metallicity of about
   0.1\,Z$_{\sun}$. We are confident with the instantaneous burst
   assumption because we are dealing with single star forming regions
   (LH). From the $B$-band total luminosity, and taking into account
   the burst strength derived, we estimate the involved total burst
   mass to be about 6.3$\times10^{4}$\,M$_{\sun}$.

   Assuming that the kinematical features observed represent the
   foreground and receding parts of the expanding bubble, the
   dynamical age for this bubble, obtained as
   $t_{\mathrm{dyn}}$=0.6$\times
   (r_{\mathrm{S}}+r_{\mathrm{N}})/(v_{\mathrm{S}}+v_{\mathrm{N}})$,
   will be about 6.5\,Myr, quite similar to the evolution time deduced
   for the starburst region, if we consider the massive stars main
   sequence time, i.e. about 4-5\,Myr, subtracting it from the
   starburst age (11.5\,Myr).

   Generally, the H$\alpha$ luminosity observed in expanding bubbles
   is accepted to be produced by photoionization from massive stars of
   the starburst precursor (MHW; Lehnert \& Heckman 1996; Martin
   1997). Then, the predicted collisionally excited H$\alpha$
   contribution should be irrelevant to the total H$\alpha$ emission
   in the expanding lobes. The H$\alpha$ luminosities measured for
   both lobes are
   L$_{\mathrm{H}\alpha}^{\mathrm{North}}$=4.7$\times10^{37}$\,erg\,s$^{-1}$
   and
   L$_{\mathrm{H}\alpha}^{\mathrm{South}}$=2.5$\times10^{38}$\,erg\,s$^{-1}$. Considering
   that both lobes are well reproduced by ellipsoids of revolution, we
   can obtain the lobe surface areas, that allow to estimate the
   colissionally produced H$\alpha$ luminosity
\begin{equation} 
        L_{\mathrm{H}\alpha} \simeq S_{\mathrm{lobe}}\ n_0\
        v_{\mathrm{exp}}\ f\ h\ \nu_{\mathrm{H}\alpha}
\label{colisionlum}
\end{equation}
   where $n_{\mathrm{0}}$ is the HI ambient density,
   $v_{\mathrm{exp}}$ the bubble expansion velocity, $f$ is the number
   of H$\alpha$ photons produced per shocked proton, and
   $\nu_{\mathrm{H}\alpha}$ is the frequency of the H$\alpha$
   photons. The predicted collisionally excited luminosities are
   L$_{\mathrm{H}\alpha}^{\mathrm{North}}$=2.4$\times10^{36}$\,erg\,s$^{-1}$
   and
   L$_{\mathrm{H}\alpha}^{\mathrm{South}}$=1.6$\times10^{36}$\,erg\,s$^{-1}$,
   less than the 10 and 1 per cent of the total H$\alpha$
   luminosity. We have assumed that approximately 0.1 H$\alpha$
   photons are produced per shocked proton. This is the value
   predicted by the Shull \& McKee \shortcite{shull79} models for a
   50\,km\,s$^{-1}$ shock. We have also adopted a HI ambient density
   of $n_{\mathrm{0}}\!\simeq$0.3 cm$^{-3}$, that is the same value
   adopted by MHW for a sample of dwarf amorphous galaxies. This value
   is obtained considering the galactocentric distance of the Mrk86--B
   bubble, 1.2\,kpc, and the galaxy HI density profile given by the
   equation~\ref{rhoHI}.

   Due to the uncertainties present in the HI total mass and scale
   measurements, we could be dealing with significantly higher HI
   ambient densities (e.g. $n_{\mathrm{0}}\!\sim$1\,cm$^{-3}$), that
   could enhance the predicted collisional H$\alpha$ luminosity. In
   any case, this contribution would not dominate the total H$\alpha$
   luminosity in the north lobe ($<30$\%), and would be negligible in
   the south lobe case ($\sim3$\%).

   Then, if the emission is photoionization dominated, the ratio
   between the luminosity of both lobes has to be related with the
   solid angle subtended from the starburst by each lobe. In fact,
   assuming revolution symmetry for both lobes, the solid angle ratio
   obtained is $\Omega_{\mathrm{South}}/\Omega_{\mathrm{North}}$=5.5
   ($\Omega_{\mathrm{South}}$=6.6\,sr and
   $\Omega_{\mathrm{North}}$=1.2\,sr), very close to the ratio
   L$_{\mathrm{H}\alpha}^{\mathrm{South}}/\mathrm{L}_{\mathrm{H}\alpha}^{\mathrm{North}}$=5.3. In
   the right panel of Figure~\ref{bubbleB} we show the aperture angles
   assumed, and the position of the ionizing radiation source (\#49
   knot in G99).
\begin{figure*}
\vspace{10.2cm}
\includegraphics{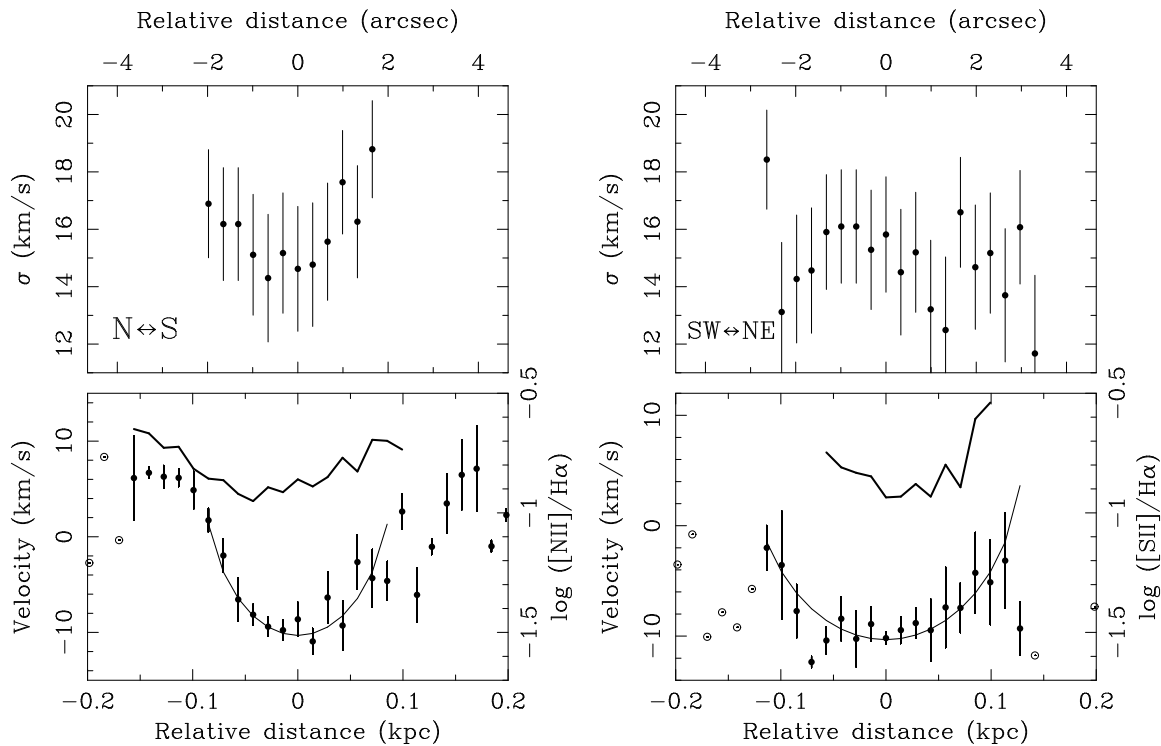}
\caption{Mrk86--C region. {\bf a)} \#7R 
and {\bf b)} \#8R long-slit velocity profiles. {\it Upper panel:}
Velocity dispersion corrected from
$\sigma_{\mathrm{instrumental}}$. {\it Lower panel:} Filled circles
represent the velocity data subtracted from large scale velocity
gradients. Dotted circles correspond to those velocities measured
using just one emission line. $\pm1\sigma$ errors bars have been
plotted. $Thin\: solid$ line represents the fit to an isotropic
expanding bubble with $v_{\mathrm{exp}}$=17.3\,km\,s$^{-1}$ and radii
90 and 130\,pc, respectively. $Thick$ lines show the [NII]/H$\alpha$
({\bf a}) and [SII]/H$\alpha$ ({\bf b}) line ratios profiles relative
to the bubble geometrical centre.}

\label{Cratios}
\end{figure*}

   In order to estimate the relevance of the shock excitation
   mechanism, we have also studied the [OII]3727\AA/[OIII]5007\AA,
   [OIII]5007\AA/H$\beta$, [NII]6583\AA/H$\alpha$ and
   ([SII]6717\AA$+$6731\AA)/H$\alpha$ line ratios across both lobes
   (see Figures~\ref{Bratios}a \&~\ref{Bratios}b). For the south lobe
   region we obtain low excitation, log([OIII]/H$\beta$)$\simeq$0.3,
   but with moderate log([OII]/[OIII])$\simeq$0.3 values. These values
   can be explained using photoionization models with low ionization
   parameter (log U$\sim$$-$4; Martin 1997). However, for the more
   remote north lobe regions the very high
   log([OII]/[OIII])$\simeq$0.6 (Figure~\ref{Bratios}a) and high
   log([NII]/H$\alpha$)$\simeq$$-$0.5 and
   log([SII]/H$\alpha$)$\simeq$0 (Figure~\ref{Bratios}b) values
   obtained imply that an {\it additional} excitation mechanism should
   be present (Lehnert \& Heckman 1996; Martin 1997). These line
   ratios are well reproduced taking into account the emission
   predicted from the models of Shull \& McKee \shortcite{shull79}
   with shock velocities of 90\,km\,s$^{-1}$. This result is
   consistent with the higher collisional H$\alpha$ luminosity
   predicted for the north lobe of the bubble.

\subsection{Mrk86--C}
\label{C}

   From the high-resolution spectra \#7R and \#8R we infer the
   presence of deep velocity minima close to the G99 \#9 knot emission
   maximum (C region in Figures~\ref{image} and~\ref{Aregion}). These
   minima observed in both spectra \#7R and \#8R (see
   Figures~\ref{Cratios}a \&~\ref{Cratios}b) are well understood if we
   are dealing with the foreground part of a highly extinct expanding
   bubble, with median parameters, $r_{\mathrm{bubble}}$=110\,pc and
   $v_{\mathrm{exp}}$=17.3\,km\,s$^{-1}$
   ($t_{\mathrm{dyn}}\!\simeq$4\,Myr). These parameters (see
   Table~\ref{parameters}) are obtained fitting the radial velocity
   profiles (global velocity gradient subtracted; $v_{\mathrm{x}}$) to
   an isotropic expanding structure velocity law (see
   Figures~\ref{Cratios}a \&~\ref{Cratios}b)
\begin{equation}  
    v_{\mathrm{x}}= v_{\mathrm{exp}}\
    \sqrt{1-\frac{\mathrm{x}^{2}}{r_{\mathrm{bubble}}^{2}}}
\label{expansion}
\end{equation} 
   The central starburst should have a high extinction, large enough
   to obscure the H$\alpha$ emission of the receding part of the
   bubble. In order to test this hypothesis we should estimate the
   physical properties, extinction included, of the starburst which
   has originated this collective wind.

   The Mrk86--C bubble has similar appearance in all the optical and
   nIR images, including the H$\alpha$ image (see
   Figure~\ref{Aregion}), in contrast with the B bubble case, probably
   due to its lower dynamical time. Its H$\alpha$ emission is slightly
   more extended along the W-E than that observed in broad-band
   filters. The corresponding optical and nIR colours measured for
   the starburst are given in the Table~\ref{parameters} and are fully
   compatible with a 7.7\,Myr old and Z=0.1\,Z$_{\sun}$ metallic
   region, adopting, in order to better fit the colours measured, an
   colour excess of E$(B-V)_{\mathrm{gas}}$=0.73.

   From these parameters, and the total $B$ absolute magnitude
   measured, applying the LH results we obtain a starburst mass of
   about 4.7$\times10^{4}$\,M$_{\sun}$. In addition, because the total
   H$\alpha$ luminosity measured for Mrk86--C,
   1.7$\times10^{39}$\,erg\,s$^{-1}$, is 6$\times10^{3}$ times higher
   than that predicted for the collisional case,
   2.7$\times10^{35}$\,erg\,s$^{-1}$, we can conclude that the line
   emission is also in Mrk86--C mainly due to photoionized gas.

   Finally, we have studied the [NII]/H$\alpha$ and [SII]/H$\alpha$
   line ratios measured from the high resolution spectra. They change
   between $\sim$0.10 in regions close to the starburst centre, and
   $\sim$0.25 and $\sim$0.30, respectively, in the external low
   H$\alpha$ surface brightness bubble regions (see
   Figures~\ref{Cratios}a \&~\ref{Cratios}b). The increase in line
   ratios values towards bubble outer regions seems again to indicate
   a decrease in the ionization parameter due to the dilution of the
   radiation from the centralized source \cite{cm97}. The values
   obtained are compatible with a pure low-metallicity HII region
   photoionization mechanism (see, e.g. Shields \& Filippenko 1990).

\section{Velocity dispersion in Mrk86--C}
\label{sigma}
   We have measured the ionized gas velocity dispersion for the \#7R
   and \#8R high-resolution spectra. From the calibration lamp and
   night sky lines we have estimated the spectral resolution in
   $\sigma_{\mathrm{instrumental}}$=16.4$\pm$2.1\,km\,s$^{-1}$. Then, we
   have obtained the velocity dispersion of the Mrk86--C bubble from
   these spectra, subtracting to $\sigma_{\mathrm{observed}}^{2}$ the
   value of $\sigma_{\mathrm{instrumental}}^{2}$. The spatial
   variation of these velocity dispersion measured is shown in
   Figure~\ref{Cratios}. The velocity dispersion obtained for the {\it
   whole} ($4\farcs6\times1\arcsec$) bubble C were
   $\sigma_{\mathrm{\#7R}}$=16$\pm$2\,km\,s$^{-1}$ and
   $\sigma_{\mathrm{\#8R}}$=15$\pm$2\,km\,s$^{-1}$. In addition, we have
   obtained the {\it mean} velocity dispersion for the whole emitting
   region, averaging the values measured in intervals of 0\farcs33,
   1\arcsec\ and 2\arcsec\ along the spatial direction. We have
   estimated 17$\pm$3 and 14.9$\pm$1.6\,km\,s$^{-1}$, respectively for
   the \#7R and \#8R spectra and 0\farcs33 intervals, 18$\pm$5 and
   15.4$\pm$0.5\,km\,s$^{-1}$ for 1\arcsec\ intervals, and finally,
   19$\pm$5 and 15.2$\pm$0.3\,km\,s$^{-1}$ for 2\arcsec\ intervals.

   We observe that there is no significant differences between local
   and total velocity dispersion values, i.e. the velocity widths
   obtained for regions of size $0\farcs33\times1\arcsec$ are similar
   to that obtained for an important fraction of the whole emitting
   region ($4\farcs6\times1\arcsec$), and that they do not change with
   the considered spatial interval$.$ 

   We have now considered the expresion for the mass of a virialized
   system given by Guzm\'an et al$.$ (1996, 1997; see also Bender,
   Burstein \& Faber 1992 and Gallego et al$.$ 1998),
\begin{equation}
   M_{\mathrm{vir}} = 1.184\ 10^{6}\ R_{\mathrm{e}} (\mathrm{kpc}) \
   \sigma^{2} (\mathrm{km/s})^{2} \;\; \mathrm{M}_{\sun} 
\label{virialmass}
\end{equation} 
   If we take for Mrk86--C the $B$-band effective radius of the
   starburst measured with {\sc cobra}, $R_{\mathrm{e}}\!\simeq$36\,pc,
   and $\sigma\!\simeq$15\,km\,s$^{-1}$, we obtain a virialized mass
   (Eq.~\ref{virialmass}) for Mrk86--C of about $M\!\sim10^{7}$\,M$_{\sun}$.

   Let us compare this result with the Mrk86--C starburst total mass
   estimated from the mass in newly formed stars
   ($M_{\mathrm{burst}}^{\mathrm{young}}$=5.3$\times10^{4}$\,M$_{\sun}$),
   underlying stellar population, gas and dark matter present in this
   region. Considering the mass density profile given in
   Sect.~\ref{global} we can estimate the mass content inside the
   effective radius of the Mrk86--C starburst precursor. Assuming the
   most favourable case, i.e. $i_{\mathrm{e}}$=0\degr\ and a dark
   matter model with $R_{\mathrm{DM}}$=2$^{1/2}\times r$, the density
   expected at a galactocentric distance, $r$=1.0\,kpc, will be about
   9\,at$_{\mathrm{H}}$\,cm$^{-3}$. Therefore, the mass content inside
   $R_{\mathrm{e}}$ should be about 10$^{5}$\,M$_{\sun}$. We
   can conclude that the burst total mass is significantly lower that
   the virialized mass deduced. The similarity of the velocity
   dispersion measured pixel by pixel rules out the global velocity
   gradient as origin for the observed velocity dispersion. First
   reason for Mrk86--C not to be bound arises from the existence of
   local gas motions induced by the Mrk86--C supernova-driven
   wind. Another argument is Mrk86--C is not placed at the galactic
   centre and, therefore, is subject of strong tidal forces that
   prevents it to be virialized.

   We have obtained the turbulent velocity dispersion
   \footnote{Turbulent velocity dispersion is defined as the line
   broadening obtained after subtracting instrumental and thermal
   broadening.} (see, e.g. Fuentes-Masip 1997), applying the expression
\begin{equation}
   \sigma_{\mathrm{turbulent}}^2 = \sigma_{\mathrm{observed}}^2 -
   \sigma_{\mathrm{thermic}}^2 - \sigma_{\mathrm{intrinsic}}^2 -
   \sigma_{\mathrm{instrumental}}^2 
\label{esigma}
\end{equation} 
   that yields a $\sigma_{\mathrm{turbulent}}\!\simeq$10.7\,km\,s$^{-1}$
   and 10.0\,km\,s$^{-1}$, respectively for \#7R and \#8R spectra,
   assuming for H$\alpha$ an
   $\sigma_{\mathrm{intrinsic}}$=7.1\,km\,s$^{-1}$ \cite{hipelein},
   and a temperature, $T$=10$^{4}$\,K. Adopting a sound velocity for
   the ionized gas in HII regions of about 12.8\,km\,s$^{-1}$
   \cite{dyson80}, the corresponding $\sigma_{\mathrm{sound}}$ is
   approximately 8\,km\,s$^{-1}$ \cite{fuentes}. Thus, we can conclude
   that the velocity dispersion in this region is sonic or slightly
   supersonic.

\section{Summary and conclusions}
\label{summary}

   \begin{enumerate}

      \item The global velocity field in Mrk86 has a central angular
      velocity of about 34\,km\,s$^{-1}$\,kpc$^{-1}$ with orientation
      PA$\sim$12\degr\ (rotation axis PA$\sim\!-$78\degr). This velocity
      gradient shows a progressive diminution towards outer galaxy
      regions, from 34\,km\,s$^{-1}$\,kpc$^{-1}$ at the galactic centre
      to $\sim$10\,km\,s$^{-1}$\,kpc$^{-1}$ at 1\,kpc. The density
      profiles of the different mass components indicate that the
      underlying stellar component dominates the total mass within its
      optical radius. The velocity gradients measured for different
      distances along the galactic rotation axis indicate that the
      ionized gas is probably distributed in a inclined rotating disk,
      with inclination in relation to the plane of the sky of about
      50\degr.

      \item High velocity gradients of about
      70\,km\,s$^{-1}$\,kpc$^{-1}$ associated with intense star
      forming regions have been observed. Both, a high mass
      concentration or a recent merger could be responsible for these
      local, steep velocity gradients.

      \item Our observations have revealed the existence of two
      bubbles, Mrk86--B and Mrk86--C. They present $v_{\mathrm{exp}}$
      of 34 and 17\,km\,s$^{-1}$ and $r_{\mathrm{bubble}}$ of 374 and
      120\,pc, respectively. These structures are driven by supernovae
      and massive stars winds originated in two low metallicity
      (Z$\sim$0.1\,Z$_{\sun}$) HII regions 11 and 8\,Myr old, with
      masses of 6.3$\times10^{4}$ and
      5.3$\times10^{4}$\,M$_{\sun}$. The H$\alpha$ emission is
      dominated by photoionization mechanisms, contributing at least
      with the 90 per cent of the total H$\alpha$
      luminosity. Moreover, the optical line ratios measured agree
      with photoionization mechanisms as the origin for their
      emission.

      \item In addition, we have studied the physical properties of
      the starburst precursor of the Mrk86--A bubble described by
      Martin \shortcite{cm98}. The optical-nIR colours measured for
      these region are well described by a 14\,Myr old, low metallicity
      (Z$\sim$0.25\,Z$_{\sun}$), 10$^{5}$\,M$_{\sun}$ massive starburst.

      The predicted starburst ages agree quite well with the dynamical
      times measured if we assume a delay of $\sim$4\,Myr between the
      starburst formation and the time when the bubble begins to
      inflate. In fact, the one tenth solar metallicity Leitherer \&
      Heckman \shortcite{leith} models predict an increment of about
      1.5dex in the deposition rate of mechanical energy 4\,Myr after
      the star formation burst.

      \item The global velocity dispersion obtained for Mrk86--C does
      not trace the total mass of the burst. The turbulent velocity
      dispersion obtained for the Mrk86--C bubble is sonic or
      slightly supersonic, with
      $\sigma_{\mathrm{turbulent}}\!\sim$10\,km\,s$^{-1}$. These values
      do not show significant changes across the region.

 \end{enumerate}

\newpage

\section*{Acknowledgements}
        Based on observations made with the Jacobus Kapteyn and Isaac
        Newton telescopes operated on the island of La Palma by the
        Royal Greenwich Observatory in the Spanish Observatorio del
        Roque de los Muchachos of the Instituto Astrof\'\i sico de
        Canarias. Based also in observations collected at the
        German-Spanish Astronomical Centre, Calar Alto, Spain,
        operated jointly by the Max-Plank-Institut f\"{u}r Astronomie
        (MPIA), Heidelberg, and the Spanish National Commission for
        Astronomy. The United Kingdom Infrared Telescope is operated
        by the Joint Astronomy Centre on behalf of the U.K. Particle
        Physics and Astronomy Council. We would like to thank
        C. S\'anchez Contreras and L. F. Miranda for obtaining the
        high resolution H$\alpha$ spectra, C. Jordi and D. Galad\'\i\,
        for obtaining the $V$ image, and A. Arag\'on-Salamanca who
        provided the $K$ image.  We also thank J. Cenarro and
        C. E. Garc\'\i a-Dab\'o for stimulating conversations. We are
        indebted to the referee, Dr. A. Burkert, for many helpful
        comments and corrections. A. Gil de Paz acknowledges the
        receipt of a {\it Formaci\'on del Profesorado Universitario}
        fellowship from the Spanish MEC. This research was supported
        by the Spanish Programa Sectorial de Promoci\'on General del
        Conocimiento under grant PB96-0610.

\bsp

\label{lastpage}


\begin{thebibliography}{} 

   \bibitem[\protect\citename{Arag\'on-Salamanca et al$.$\
   }1993]{alfonso} Arag\'on-Salamanca A., Ellis R. S., Couch W. J.,
   Carter D., 1993, MNRAS, 262, 764

   \bibitem[\protect\citename{Arp\ }1966]{arp66} Arp H. 1966, Atlas of
   Peculiar Galaxies, California Institute of Technology, Pasadena

   \bibitem[\protect\citename{Bender, Burstein \& Faber\
   }1992]{bender} Bender R., Burstein D., Faber S. M., 1992, ApJ, 399,
   462 
   
   \bibitem[\protect\citename{Binney \& Tremaine\ }1987]{binney}
   Binney J.,Tremain S., 1987, in Galactic Dynamics,
   ed. J. P. Ostriker, Princeton University Press

   \bibitem[\protect\citename{Blok et al$.$\ }1996]{blok} de Blok
   W. J. G., McGaugh s. s., van der Hulst J. M., 1996, MNRAS, 283, 18

   \bibitem[\protect\citename{Bomans, Chu \& Hopp\ }1997]{bomans} Bomans 
   D. J., Chu Y.-H., Hopp U., 1997, AJ, 113, 1678

   \bibitem[\protect\citename{Bottinelli et al$.$\ }1984]{botti}
   Bottinelli L., Gouguenheim L., Paturel G., de Vaucouleurs G., 1984,
   A\&AS, 56, 381

   \bibitem[\protect\citename{Brinks\ }1994]{brinks} Brinks E., 1994,
   in Violent Star Formation: From 30 Doradus to QSOs, ed.\
   G. Tenorio-Tagle, 1st. IAC-RGO meeting, La Palma, p. \ 145

   \bibitem[\protect\citename{Broeils\ }1992]{broeils} Broeils A. H.,
   1992, Ph. D. Thesis, Rijksuniversiteit Groningen

   \bibitem[\protect\citename{Bruzual \& Charlot \ }1996]{bc96}
   Bruzual G., Charlot S., 1996, unpublished

   \bibitem[\protect\citename{Burkert\ }1995]{burkert} Burkert A.,
   1995, ApJ, 447, 25

   \bibitem[\protect\citename{Calzetti\ }1997]{calze} Calzetti D.,
   1997, to appear in The Ultraviolet Universe at Low and High
   Redshift, preprint
 
   \bibitem[\protect\citename{Carignan \& Beaulieu\ }1989]{carignan} Carignan 
   C., Beaulieu S., 1989, ApJ, 347, 760

   \bibitem[\protect\citename{Carignan \& Freeman\ }1988]{cf} Carignan
   C., Freeman K. C., 1988, ApJ, 332, L33

   \bibitem[\protect\citename{Castor et al$.$\ }1975]{castor} Castor J.,
     McCray R., Weaver R., 1975, ApJ, 200, L107

   \bibitem[\protect\citename{Chevalier \& Clegg\ }1985]{cheva}
   Chevalier R., Clegg A., 1985, Nature, 317, 44

   \bibitem[\protect\citename{De Vaucouleurs \& Pence\ }1980]{vauco}
   De Vaucouleurs G., Pence W. D., 1980, ApJ, 242, 18

   \bibitem[\protect\citename{De Young \& Gallagher\ }1990]{deyoung}
   De Young D. S., Gallagher J. S., 1990, ApJ, 356L, 15

   \bibitem[\protect\citename{Dultzin-Hacyan et al$.$\ }1990]{dultzin}
   Dultzin-Hacyan D., Masegosa J., Moles M., 1990, A\&A, 238, 28

   \bibitem[\protect\citename{Dyson \& Williams\ }1980]{dyson80} Dyson
   J. E., Williams D. A., 1980, in The Physics of the Interstellar Medium
   (Manchester: Manchester University Press)

   \bibitem[\protect\citename{Fanelli et al$.$\ }1988]{fan:con} Fanelli
   M.N., O'Conell R.W., Thuan T.X., 1988, ApJ, 334, 665

   \bibitem[\protect\citename{Fernie\ }1983]{fernie} Fernie J. D.,
   1983, PASP, 95, 782 

   \bibitem[\protect\citename{Flores \& Primack\ }1994]{flores} Flores
   R. A., Primack J. R., 1994, ApJ, 427, L1

   \bibitem[\protect\citename{Fuentes-Massip\ }1997]{fuentes}
   Fuentes-Masip O., 1997, Ph. D. Thesis, Universidad de La Laguna

   \bibitem[\protect\citename{Gallego et al$.$\ }1998]{gallego} Gallego
   J., Zamorano J., Garc\'{\i}a-Dab\'o C. E., Arag\'on-Salamanca A.,
   1998, in The Young Universe, ASP conference series, 146, 235

   \bibitem[\protect\citename{Gil de Paz et al$.$\ }1998]{gilm} Gil de
   Paz A., Zamorano J., Gallego J., 1998, in Dwarf Galaxies and
   Cosmology. XVIII Reencontres de Moriond, eds: T.V., Thuan T. X.,
   Balkowski C., Cayatte V., and Van T. T., Les Arcs. (G98)

   \bibitem[\protect\citename{Gil de Paz et al$.$\ }1999]{gil} Gil de
   Paz A., Zamorano J., Gallego J., Arag\'on-Salamanca A., Dom\'\i
   nguez, F. de B., 1999, in prep. (G99)

   \bibitem[\protect\citename{Guzm\'an et al$.$\ }1996]{guzman2} Guzm\'an
   R., Koo D. C., Faber S. M., et al$.$, 1996, ApJ, 460, L5

   \bibitem[\protect\citename{Guzm\'an et al$.$\ }1997]{guzman} Guzm\'an
   R., Gallego J., Koo D. C., et al$.$, 1997, ApJ, 489, 559

   \bibitem[\protect\citename{Heckman et al$.$\ }1990]{heck90} Heckman
   T. M., Armus L., Miley G. K., 1990, ApJS, 74, 833 

   \bibitem[\protect\citename{Hippelein\ }1986]{hipelein} Hippelein
   H., 1986, A\&A, 160, 374
    
   \bibitem[\protect\citename{Hodge \& Kennicutt\ }1983]{hodge} Hodge
   , Kennicutt R. C., 1983, ApJ, 265, 132

   \bibitem[\protect\citename{Israel et al$.$\ }1995]{israel} Israel 
   F. P., Tacconi L. J., Baas F., 1995, A\&A, 295, 599 

   \bibitem[\protect\citename{Izotov et al$.$\ }1996]{izotov} Izotov
   Y. I., Dyak A. B., Chaffee F. H., et al$.$, 1996, ApJ, 458, 524.

   \bibitem[\protect\citename{Kamphuis et al$.$\ }1996]{kamph} Kamphuis
   J. J., Sijbring L. G., van Albada T. S., 1996, A\&AS, 116, 15

   \bibitem[\protect\citename{Kent\ }1985]{kent} Kent S. M., 1985,
   PASP, 97, 165

   \bibitem[\protect\citename{Klein et al$.$\ }1984]{kl84} Klein U.,
   Wielebinski R., Thuan T.X., 1984, A\&A, 141, 241

   \bibitem[\protect\citename{Klein et al$.$\ }1991]{kl91} Klein U.,
   Weiland H., Brinks E., 1991, A\&A, 246, 323

   \bibitem[\protect\citename{Leitherer \& Heckman\ }1995]{leith}
   Leitherer C., Heckman T. M., 1995, ApJS, 96, 9 (LH)

   \bibitem[\protect\citename{Lehnert \& Heckman\ }1996]{lehnert}
   Lehnert M. D., Heckman T. M., 1996, ApJ, 462, 651

   \bibitem[\protect\citename{Lo et al$.$\ }1993]{lo} 
   Lo K. Y., Sargent W. L. W., Young K., 1993, AJ, 106, 507

   \bibitem[\protect\citename{Longo et al$.$\ }1991]{lon:cap} Longo G.,
   Capaccioli M, Ceriello A., 1991, A\&AS, 90, 375

   \bibitem[\protect\citename{Lonsdale et al$.$\ }1985]{lonsdale}
   Lonsdale C.J., Helou G., Good J.C., Rice W., 1985, Catalogued
   Galaxies and Quasars Observed in the IRAS Survey. Jet Propulsion
   Lab., Pasadena.

   \bibitem[\protect\citename{Loose \& Thuan\ }1985]{loose} Loose
   H.-H., Thuan T.X., 1985, in Star-Forming Dwarf Galaxies, eds. Kunth
   D., Thuan T.X., and Van J.T.T. Editions Fronti\`{e}res.

   \bibitem[\protect\citename{Loose \& Thuan\ }1986]{loose86} Loose
   H.-H., Thuan T.X., 1986, ApJ, 309, 59

   \bibitem[\protect\citename{Mac Low \& Ferrara\ }1998]{mclow} Mac Low
   M.-M., Ferrara A., 1998, preprint, astro-ph/9801237 

   \bibitem[\protect\citename{Maeder\ }1990]{maeder} Maeder A., 1990,
   A\&AS, 84, 139

   \bibitem[\protect\citename{Markarian\ }1969]{mark69} Markarian
   B. E., 1969, Afz, 5, 443
  
   \bibitem[\protect\citename{Marlowe et al$.$\ }1995]{marlowe} Marlowe
   A., Heckman T. M., Wyse R. F. G., Schommer R., 1995, ApJ, 438, 563
   (MHW)

   \bibitem[\protect\citename{Martin\ }1996]{cm96} Martin C. L., 1996,
   ApJ, 465, 680

   \bibitem[\protect\citename{Martin\ }1997]{cm97} Martin C. L., 1997,
   ApJ, 491, 561

   \bibitem[\protect\citename{Martin\ }1998]{cm98} Martin C. L., 1998,
   ApJ, preprint, astro-ph/9804165 (CM98)

   \bibitem[\protect\citename{Mathis\ }1990]{mathis} Mathis J. S.,
   1990, ARA\&A, 28, 37
  
   \bibitem[\protect\citename{Meurer et al$.$\ }1998]{meurer} Meurer
   G. R., Staveley-Smith L., Killeen N. E. B., 1998, MNRAS, preprint,
   astro-ph/9806261

   \bibitem[\protect\citename{Moore\ }1994]{moore} Moore B., 1994,
   Nature, 370, 629

   \bibitem[\protect\citename{Mori et al$.$ \ }1997]{mori} Mori M.,
   Yuzuru Y., Takuji T., Ken'ichi N., 1997, ApJ, 478, L21

   \bibitem[\protect\citename{Navarro, Eke \& Frenk\ }1996]{nef}
   Navarro J. F., Eke V. R., Frenk C. S., 1996, MNRAS, 283, 72

   \bibitem[\protect\citename{Navarro, Frenk \& White\ }1997]{nfw}
   Navarro J. F., Frenk C. S., White S. D. M., 1997, ApJ, 490, 493

   \bibitem[\protect\citename{Ojha \& Joshi\ }1991]{oj:jos} Ojha D.K.,
   Joshi S.C., 1991, Ap\&SS, 183, 245

   \bibitem[\protect\citename{\"Ostlin et al$.$\ }1998]{ostlin} \"Ostlin
   G., Amram P., Masegosa J., Bergvall N., Boulesteix J., 1998,
   preprint, astro-ph/9812283

   \bibitem[\protect\citename{Papaderos et al$.$\ }1996a]{papa1}
   Papaderos P., Loose H.-H., Thuan T. X., Fricke K. J., 1996a, A\&AS,
   120, 207 (P96a)

   \bibitem[\protect\citename{Papaderos et al$.$\ }1996b]{papa2}
   Papaderos P., Loose H.-H., Fricke K. J., Thuan T. X., 1996b, A\&A,
   314, 59 (P96b)

   \bibitem[\protect\citename{Petrosian et al$.$\ }1997]{petro}
   Petrosian A. R., Boulesteix J., Comte G., Kunth D., LeCoarer E.,
   1997, A\&A 318, 390

   \bibitem[\protect\citename{Pryor \& Kormendy\ }1990]{pryor} Pryor
   C., Kormendy J., 1990, AJ, 100, 127

   \bibitem[\protect\citename{Puche et al$.$\ }1992]{puche} Puche D., 
   Westphal D., Brinks E., Roy J. R., 1992, AJ, 103, 1841

   \bibitem[\protect\citename{Roy et al$.$\ }1991]{roy} Roy J. R.,
   Boulesteix J., Joncas J., Grundseth B., 1991, ApJ, 367, 141

   \bibitem[\protect\citename{Sage et al$.$\ }1992]{sage} Sage L. J.,
   Salzer J. J., Loose H.-H., Henkel C., 1992, A\&A, 265, 19

   \bibitem[\protect\citename{Sait\={o} et al$.$\ }1992]{saito} Sait\={o}
   M., Sasaki M., Ohta K., Yamada T., 1992, PASJ, 44, 593

   \bibitem[\protect\citename{Salpeter \ }1955]{salpeter} Salpeter,
   E. E., 1955, ApJ, 121, 161

   \bibitem[\protect\citename{Salucci \& Persic\ }1997]{salucci}
   Salucci P., Persic M., 1997, in Dark and Visible Matter in
   Galaxies, eds. M. Persic \& P. Salucci, ASP conference series, 117,

   \bibitem[\protect\citename{Scalo\ }1986]{scalo} Scalo J. M., 1986,
   Fund. Cosmic Phys., 11, 1

   \bibitem[\protect\citename{Shapley \& Ames\ }1932]{shap32} Shapley
   H., Ames A., 1932, in Annals of the Harvard College Obs. 88, No. 2

   \bibitem[\protect\citename{Shields \& Filippenko\ }1990]{shields}
   Shields G. A., Filippenko A. V., 1990, AJ, 100, 103

   \bibitem[\protect\citename{Shull & McKee\ }1979]{shull79} Shull
   S. M., McKee C., 1979, ApJ, 227, 131

   \bibitem[\protect\citename{Silk, Wyse \& Shields\ }1987]{silk} Silk
   J., Wyse R. F. G., Shields G., 1987, ApJ, 322, L59

   \bibitem[\protect\citename{Skillman \& Kennicutt\ }1993]{skill93}
   Skillman E. D., Kennicutt R. C., 1993, ApJ, 411, 655
   
   \bibitem[\protect\citename{Staveley-Smith et al$.$\ }1992]{ss}
   Staveley-Smith L., Davies R. D., Kinman T. D., 1992, MNRAS, 258, 334

   \bibitem[\protect\citename{Swaters\ }1998]{swaters} Swaters R.,
   1998, in Galaxy Dynamics, Rutgers University, preprint,
   astro-ph/9811010

   \bibitem[\protect\citename{Thronson et al$.$\ }1989]{thr} Thronson
   H., Tacconi L., Kenney J., et al$.$, 1989, ApJ, 344, 747

   \bibitem[\protect\citename{Thuan \ }1983]{thu83} Thuan
   T. X., 1983, ApJ, 268, 667

   \bibitem[\protect\citename{Thuan\ }1991]{thu91} Thuan T. X., 1991,
      in Massive Stars in Starburst, eds.\ C. Leitherer,
      N. R. Walborn, T. M. Heckman, C. A. Norman, StSci Symposium
      series 5, Baltimore, p. \ 183

   \bibitem[\protect\citename{Thuan \& Martin\ }1981]{thu81} Thuan
   T. X., Martin G. E., 1981, ApJ, 247, 823

   \bibitem[\protect\citename{Tomita et al$.$\ }1997]{tomi} Tomita A.,
   Ohta K., Nakanishi K., Takeuchi T. T., Sait\={o} M., 1997, AJ, 116,
   131

   \bibitem[\protect\citename{Vader\ }1986]{vader86} Vader J. P.,
   1986, ApJ, 305, 669

   \bibitem[\protect\citename{Vader\ }1987]{vader87} Vader J. P.,
   1987, ApJ, 317, 128

   \bibitem[\protect\citename{Van Zee et al$.$\ }1998]{vanzee} Van Zee L., Skillman E. D., Salzer J. J., 1998, preprint, astro-ph/9806246

   \bibitem[\protect\citename{Verter\ }1985]{verter} Verter F., 1985,
   ApJS, 57, 261

   \bibitem[\protect\citename{Young \& Knezek\ }1989]{yk} Young J. S.,
   Knezek P., 1989, ApJ, 347, L55

\end{thebibliography}
\end{document}